\documentclass[journal]{IEEEtran}

\usepackage{amsmath,amssymb,amsfonts,amsthm}
\usepackage{algorithm,algorithmic}
\usepackage{graphicx}
\usepackage{textcomp}
\usepackage{xcolor}
\usepackage{booktabs}
\usepackage{multirow}
\usepackage{adjustbox}   %
\usepackage{cite}
\usepackage[hidelinks]{hyperref}
\usepackage{url}
\usepackage{array}
\usepackage{tikz}
\usepackage{fontawesome5}   %
\usetikzlibrary{positioning,arrows.meta,calc,backgrounds,shadows.blur}
\allowdisplaybreaks

\definecolor{cUAVe}{HTML}{1F8A8A}
\definecolor{cHAPS}{HTML}{1F4E79}

\title{AeroGLIDER: Hierarchical Graph Reinforcement Learning for Joint Task Offloading and Resource Allocation in HAPS-UAV-IoT Networks\thanks{This work has been submitted to the IEEE for possible publication. Copyright may be transferred without notice, after which this version may no longer be accessible.}}

\author{
\IEEEauthorblockN{Xuli Cai,~\IEEEmembership{Student~Member,~IEEE}, Poonam Lohan,~\IEEEmembership{Senior~Member,~IEEE}, Burak Kantarci,~\IEEEmembership{Fellow,~IEEE}}\\
\IEEEauthorblockA{\textit{University of Ottawa, Ottawa, ON, Canada},
\{xcai049, ppoonam, burak.kantarci\}@uottawa.ca}
}

\begin{document}
\maketitle

\begin{abstract}
High-altitude platform station (HAPS)-assisted edge computing enables Internet of Things (IoT) tasks to be processed across three tiers: local IoT devices, low-altitude uncrewed aerial vehicles (UAVs), and the HAPS. This paper studies joint task offloading and resource allocation in such HAPS-UAV-IoT systems, aiming to maximize the fraction of tasks completed within their deadlines under an explicit energy budget. Many existing deep reinforcement learning (DRL) approaches rely on fixed-length state and action representations, which makes them difficult to apply when the number of active tasks or devices changes over time. We propose AeroGLIDER, a hierarchical graph reinforcement learning framework that represents the system as a heterogeneous task-resource graph and makes decisions at two timescales: a slow manager makes offloading decisions, while a fast worker allocates communication and computation resources to each task. The worker generates actions on task-resource graph edges rather than as a fixed-size vector, so the action space grows or shrinks with the workload, and a single trained policy operates under varying task arrivals and network sizes without retraining. Congestion-aware features (e.g., queue depth, backlog) and a relational critic improve decisions under heavy traffic. Simulations across multiple task arrival rates and deployment configurations show that AeroGLIDER achieves the highest task success rate at every arrival rate, with a margin over the strongest baseline that widens from $1.2$ to $13.9$ percentage points as congestion increases. AeroGLIDER also generalizes zero-shot to arrival rates up to 50\% beyond training and to network sizes from half to twice the training scale.
\end{abstract}

\section{Introduction}\label{sec:introduction}
\IEEEPARstart{H}{igh-altitude} platform stations (HAPSs) are emerging as an important extension of mobile edge computing (MEC). Operating in the stratosphere at about $20$~km altitude, a HAPS provides wide line-of-sight coverage and onboard computation that complement ground Internet of Things (IoT) devices and uncrewed aerial vehicles (UAVs) at low altitude~\cite{kurt2021hapsmag, lou2023hapsedge}. The result is a three-tier architecture in which an IoT task can be processed locally, offloaded to a nearby UAV, or forwarded to the HAPS. Resource management in this architecture is demanding: tasks differ in input size, computation demand, and deadline; the tiers differ in communication (comm.), computation, and energy constraints; and the number of active tasks and the congestion at each tier change continuously. Task offloading and resource allocation must therefore be decided jointly and adaptively.
\subsection{Motivation and Limitations of Prior Work}
Deep reinforcement learning (DRL) is widely used for joint task offloading and resource allocation in MEC~\cite{wang2020drlsurvey}, but three common design choices limit its use in dynamic HAPS-UAV-IoT networks. First, most methods flatten the network state into a vector of fixed length, sized for a predefined number of devices or tasks~\cite{chen2019offloading}. When the number of active tasks changes, the policy must drop the excess tasks or pad the state. This ties the policy to one network size and handles bursty arrivals poorly. Second, methods that train a separate learner for each task or device~\cite{tang2022offloading} keep each decision small but hide the current shared resource state. Tasks then compete for the same transmitter, backhaul link, or compute unit without a common view of congestion. Third, recent DRL methods built on graph neural networks (GNNs) capture the network topology, but they pool node embeddings into one vector for the whole graph before acting~\cite{wu2025dependency}. This dilutes the signals of individual resources, such as queue depth and backlog, and makes it hard to tell which task-resource decision caused a given outcome. A scalable policy should instead avoid action vectors of fixed size, keep the relations between tasks and resources, and act on local congestion information.
\subsection{Proposed Approach and Contributions}
We propose AeroGLIDER (Aero Graph Learning for Intelligent Distributed Edge Resource Allocation), a hierarchical graph reinforcement learning framework for HAPS-assisted edge computing. AeroGLIDER models the active tasks, comm. links, and computing resources as a heterogeneous task-resource graph and operates at two timescales: a manager splits each task across the three tiers once per macro-slot, and a worker allocates bandwidth, transmit power, and CPU once per micro-slot. The manager decides from the graph relations between each task and the computing tiers, and the worker outputs its actions directly on the task-resource edges. The decision interface therefore grows and shrinks with the current workload instead of being sized for a maximum task count.

AeroGLIDER builds on our earlier HALO framework~\cite{cai2026prior}, which combined an auction for device-UAV association with a two-timescale hierarchical PPO controller for the same three-tier system. HALO, however, uses flat state and action vectors over $N_{\max}$ task slots padded with zeros, observes only the total UAV and HAPS backlogs, and was evaluated on a single deployment over a narrow traffic range without an energy constraint. This paper keeps HALO's two-timescale structure and admission auction and replaces its learning core: the flat vectors become a heterogeneous task-resource graph with actions on its edges, congestion features for each resource, and a relational critic, and an energy budget per macro-slot is enforced through a Lagrangian constraint. The evaluation is broadened to seven arrival rates, five deployment configurations, a model-based reference, ablations, and zero-shot tests. Specifically, this paper makes the following contributions:
\begin{itemize}
  \item \textit{Resource allocation policy on the edges of a heterogeneous graph}: The manager makes an offloading decision for each task from the resource states encoded in the graph, and the worker generates comm. and computation allocations directly on the task-resource edges. The worker action therefore has no fixed dimension, the interface follows the number of active tasks without padding, truncation, or retraining, and the policy does not depend on task ordering. A softmax over the tasks of each resource guarantees that the bandwidth and computing allocations respect the capacity constraints.
  \item \textit{Congestion-aware graph design with a relational critic}: Backlog and queue depth are attached to every resource node, slack and urgency to every task node, and the critic pools information from the resources involved in each decision rather than from the whole graph. This improves value estimation under heavy traffic and helps the policy reason about resource contention.
  \item \textit{Comprehensive evaluation against learned and model-based references}: Across seven arrival rates and five deployment configurations, AeroGLIDER achieves the highest task success rate among the DRL methods (PPO, SAC, and DDPG) at every arrival rate, and stays within $4.6$-$5.1$ percentage points of a receding horizon linear programming (RH-LP) reference. The same trained policy generalizes zero-shot to arrival rates up to $50\%$ beyond training and to network sizes from half to twice the training scale.
\end{itemize}
The rest of this paper is organized as follows. Section~\ref{sec:related} reviews the related literature. Section~\ref{sec:system_model} introduces the system model and the problem formulation. Section~\ref{sec:framework} presents the AeroGLIDER framework. Section~\ref{sec:results} reports the simulation results, and Section~\ref{sec:conclusion} concludes the paper.

\section{Related Work}
\label{sec:related}
We position AeroGLIDER against three lines of work: (i) DRL for task offloading in MEC and UAV networks, (ii) HAPS-assisted edge computing, and (iii) DRL with GNNs for resource allocation.
\subsection{DRL for Task Offloading in MEC and UAV Networks}
Early work cast offloading as a Markov decision process (MDP) and solved it with deep Q-networks \cite{mnih2015dqn} and offloading policies over discrete decisions \cite{huang2019deepoffload, chen2018drloffloading}. Later work adopted policy gradient and actor-critic methods (PPO \cite{schulman2017ppo}, DDPG \cite{lillicrap2016ddpg}, and SAC \cite{haarnoja2018sac}) for continuous bandwidth and power allocation; other work extended offloading to UAV-assisted MEC that serves ground IoT devices \cite{cheng2019uavmec, zhang2020uavmec}. Many of these methods treat the state as a flat vector \cite{chen2019offloading, min2019offloading} sized for a fixed device count.
Hierarchical reinforcement learning (RL) splits control across time scales \cite{vezhnevets2017feudal, nachum2018hierarchical} and has been applied to computation offloading \cite{zhou2023hierarchical}, where both control levels still operate on flat state vectors. We keep the two-timescale decomposition but replace the flat states with heterogeneous graphs.

\subsection{HAPS-Assisted Edge Computing}
The broader setting places the HAPS within space-air-ground integrated networks (SAGIN), which extend terrestrial cellular and edge computing infrastructure into the stratosphere \cite{kurt2021hapsmag, lou2023hapsedge}. For computation offloading, Wu et~al.~\cite{wuMultiHAPAssistedComputationOffloading2025} studied offloading to several HAPSs with a latency objective; Jia et~al.~\cite{jiaHierarchicalAerialComputing2023} and Nabi and Moh~\cite{nabiJointOffloadingDecision2025} shared HAPS and UAV resources among IoT devices through matching games; and a parallel line of work solves SAGIN offloading with multi-agent, federated, or perception-aware DRL and with game theory \cite{wangComputationOffloadingMultiAgent2024, yangTaskPredictionBasedEdge2025, tangBlockchainBasedTrustedTraffic2022, waqasPerceptionAwareOffloadingCollaborative2025, zhangEvolutionaryGameResource2025}. Closest to our setting, queue-aware HAPS-UAV-IoT offloading has been addressed with multi-agent DRL~\cite{hevesli2025queueaware}, and hierarchical multi-agent RL (MARL) with auction-assisted coordination has been studied for low-altitude networks~\cite{you2025auctionlain}. These approaches, however, keep a fixed decision interface or one learner per agent; none acts on a task-resource action space whose size follows the workload.

\begin{table*}[!tp]
\centering
\caption{Positioning of AeroGLIDER Related to Representative Prior Work.}
\label{tab:positioning}
\renewcommand{\arraystretch}{1.08}
\footnotesize
\adjustbox{max width=\textwidth}{%
\begin{tabular}{l l c c c c c}
\toprule
\textbf{Method} & \textbf{Year} &
\textbf{State repr.} &
\textbf{Action gran.} &
\textbf{Hierarchical} &
\textbf{Congestion-aware} &
\textbf{HAPS-UAV-IoT} \\
\midrule
DROO \cite{huang2019deepoffload}       & 2020 & Flat vector  & Per-device     & No  & No  & No \\
H-MADRL \cite{zhou2023hierarchical}      & 2023 & Flat vector  & Per-device     & Yes & No      & No \\
GNN-DRL \cite{sun2023graphrl}      & 2023 & Homog. GNN   & Edge-level     & No  & Partial & No \\
HGNN-MEC \cite{wu2025dependency}    & 2025 & Heterog. GNN & Pair-level     & No  & Partial & No \\
HAPS-MEC \cite{wuMultiHAPAssistedComputationOffloading2025}& 2025 & MINLP opt.   & Analytic       & No  & No  & No \\
HALO (our prior work) \cite{cai2026prior} & 2026 & Flat vector  & Per-task slot  & Yes & Partial & Yes \\
\textbf{AeroGLIDER (ours)}                & --- & \textbf{Heterog. GNN} & \textbf{Edge-level} & \textbf{Yes} & \textbf{Yes} & \textbf{Yes} \\
\bottomrule
\multicolumn{7}{@{}p{0.96\textwidth}@{}}{\footnotesize
``Action gran.'' denotes the level at which control decisions are produced
(per-device, per-task slot of a fixed-length vector, pair-level, or edge-level).
``Analytic'' indicates a scheduler based on optimization rather than on learning.
For congestion awareness, \textbf{Yes} denotes the use of
per-resource queue depth and backlog, \textbf{Partial} denotes the
use of aggregate congestion indicators only, and \textbf{No} denotes the
absence of congestion information. AeroGLIDER explicitly incorporates
per-resource queue depth and backlog in its graph state
(Sec.~\ref{sec:graphs}).}
\end{tabular}}
\end{table*}
\subsection{Graph Neural Networks for Resource Allocation}
GNNs have become an important tool for wireless and edge resource allocation because message passing between nodes naturally captures the interactions among devices, links, and shared resources \cite{eisen2020gnnwireless, shen2021gnnradio}. Common backbones include graph convolutional networks (GCN) \cite{kipf2017gcn}, graph attention networks (GAT) \cite{velickovic2018gat}, graph isomorphism networks (GIN) \cite{xu2019gin}, and the inductive GraphSAGE aggregation scheme \cite{hamilton2017graphsage}. Heterogeneous graph extensions further support systems with several node and relation types \cite{schlichtkrull2018rgcn, wang2019han}. Closer to computation offloading, recent DRL methods with GNN encoders \cite{sun2023graphrl, wu2025dependency} use graph representations to improve topology awareness and scalability. Their decision interfaces, however, are coarser than the task-resource edges used here: they either pool the node embeddings into one vector for the whole graph and then select one node pair per step~\cite{wu2025dependency}, or classify each device-AP edge as a binary decision~\cite{sun2023graphrl}. Graph RL has also been studied outside MEC, for vertex selection in combinatorial optimization \cite{khalil2017comb}, job shop dispatching \cite{zhang2020l2d}, and the scheduling of DAG stages \cite{mao2019decima}. These formulations typically select one discrete node or operation per decision step with a single policy level.

AeroGLIDER differs in that the worker produces continuous allocations directly on the task-resource edges, without a separate matching stage, and keeps the allocations of each resource within its capacity. Actions on graph edges are not new in themselves. The contribution lies in combining them with a relational critic, congestion-aware features, and two-timescale hierarchical control for joint offloading and resource allocation in HAPS-UAV-IoT systems, a combination that, to our knowledge, has not been explored before.

Table~\ref{tab:positioning} highlights the main distinction:
existing graph methods act at edge or pair level, whereas HALO uses a hierarchical controller with a fixed task-slot interface. AeroGLIDER combines heterogeneous graph states, task-resource edge actions, two-timescale control, and per-resource congestion
information.

\begin{table*}[!tp]
\centering
\caption{Summary of Notations}
\label{tab:notation}
\renewcommand{\arraystretch}{0.92}
\scriptsize
\adjustbox{max width=\textwidth}{%
\begin{tabular}{@{}l@{\hskip 3pt}p{0.225\textwidth}@{\hskip 3pt}l@{\hskip 10pt}l@{\hskip 3pt}p{0.225\textwidth}@{\hskip 3pt}l@{}}
\toprule
\textbf{Symbol} & \textbf{Description} & \textbf{Ref.} & \textbf{Symbol} & \textbf{Description} & \textbf{Ref.} \\
\midrule
\multicolumn{6}{@{}l}{Sets, indices, and time scales}\\
$\mathcal{N},N$ & set / number of IoT devices & Sec.~\ref{sec:system_model} & $\mathcal{M},M$ & set / number of UAVs & \eqref{eq:feasible_uav} \\
$\mathcal{N}_{\mathrm{all}}$ & set of all tasks over the horizon & \eqref{eq:opt_obj} & $\mathbb{L}$ & training set of task arrival rates & \eqref{eq:training_loads}, Sec.~\ref{sec:train} \\
$t,k,\tau$ & macro-slot / micro-slot within a macro-slot / global micro-slot index & Sec.~\ref{sec:traffic} & $\Delta t,\Delta T,K_{\mathrm{slots}}$ & micro-slot / macro-slot duration, micro-slots per macro-slot ($\Delta T{=}K_{\mathrm{slots}}\Delta t$) & Sec.~\ref{sec:traffic} \\
$T$ & episode length (macro-slots) & \eqref{eq:avg_energy} &  &  &  \\
\midrule
\multicolumn{6}{@{}l}{Task}\\
$\phi_i$ & task $i$: $(D_i,C_i,T_i^{\max},\tau_i^{\mathrm{arr}})$ & \eqref{eq:task_model} & $D_i,C_i$ & input size (bits), compute intensity & \eqref{eq:task_model} \\
$T_i^{\max},\tau_i^{\mathrm{arr}}$ & latency deadline, arrival micro-slot & \eqref{eq:task_model} & $\alpha_i,\beta_i,\gamma_i$ & local / edge / cloud split ($\sum{=}1$) & \eqref{eq:task_split} \\
$\ell,W_i^{\ell}$ & phase $\ell\!\in\!\{$a2g,bh,loc,uav,haps$\}$; per-phase workload & \eqref{eq:phase}, \eqref{eq:phase_workloads} & $S_i^{\ell}[\tau],\tau_i^{\ell,\mathrm{start}}$ & cumulative service / start slot of phase $\ell$ & \eqref{eq:cumulative_service}, \eqref{eq:phase_dependency} \\
$\tau_i^{\ell,\mathrm{fin}}$ & phase $\ell$ completion (global micro-slot) & \eqref{eq:phase_completion} & $\Lambda$ & aggregate task arrival rate (Poisson) & Sec.~\ref{sec:traffic}, \eqref{eq:training_loads} \\
$s_i,u_i$ & slack feature, urgency level & \eqref{eq:slack_ratio}, Sec.~\ref{sec:association} & $\mathcal{M}_i,d_c$ & task $i$'s feasible UAV set, auction distance scale & \eqref{eq:feasible_uav}, Sec.~\ref{sec:association} \\
\midrule
\multicolumn{6}{@{}l}{Channel}\\
$\theta_{i,m}$ & elevation angle (IoT$\,i\to$UAV$\,m$) & \eqref{eq:plos} & $H_m,H_{\mathrm{H}}$ & UAV / HAPS altitude & Sec.~\ref{sec:association}, Tab.~\ref{tab:sim_params} \\
$d_{i,m},d_{m,\mathrm{H}}$ & IoT-UAV / UAV-HAPS distance & \eqref{eq:pathloss}, \eqref{eq:a2a_gain} & $d_{i,m}^{\mathrm{2D}},R_{\mathrm{cov}}$ & horizontal IoT-UAV distance, UAV coverage radius & \eqref{eq:feasible_uav} \\
$P_{\mathrm{LoS}},a,b$ & LoS probability and its model parameters & \eqref{eq:plos} & $\overline{L}_{i,m},L_{\mathrm{FS}}$ & mean / free-space path loss & \eqref{eq:pathloss} \\
$\eta_{\mathrm{LoS}},\eta_{\mathrm{NLoS}}$ & LoS / NLoS excess attenuation & \eqref{eq:pathloss} & $h_{i,m},h_{m,\mathrm{H}}$ & A2G / A2A channel gains & \eqref{eq:a2g_gain}, \eqref{eq:a2a_gain} \\
$\rho_0,\alpha_{\mathrm{AA}},G_{\mathrm{A2A}}$ & ref.\ gain, A2A exponent, A2A antenna gain & \eqref{eq:a2a_gain} & $f_c,c$ & carrier frequency, speed of light & \eqref{eq:pathloss} \\
$N_0,\Gamma_{i,m}^{\mathrm{U}},\Gamma_{i,m}^{\mathrm{H}}$ & noise PSD, access / backhaul SINR & \eqref{eq:a2g_sinr}, \eqref{eq:haps_sinr} & $I_{i,m}$ & inter-cell interference power & \eqref{eq:total_interference} \\
\midrule
\multicolumn{6}{@{}l}{Comm. and computation}\\
$W_{\mathrm{IU}},W_{\mathrm{UH}}$ & IoT$\to$UAV / UAV$\to$HAPS bandwidth & \eqref{eq:a2g_rate}, \eqref{eq:haps_rate} & $w_i^{\mathrm{U}},w_i^{\mathrm{H}}$ & allocated bandwidth fractions & \eqref{eq:a2g_rate}, \eqref{eq:haps_rate} \\
$p_i^{\mathrm{IoT}},p_{i,m}^{\mathrm{UAV}}$ & allocated transmit powers & \eqref{eq:a2g_sinr}, \eqref{eq:haps_sinr} & $P_{\max}^{\mathrm{IoT}},P_{\max}^{\mathrm{UAV}}$ & maximum transmit power & \eqref{eq:opt_iot_power}, \eqref{eq:opt_uav_power} \\
$R_{i,m}^{\mathrm{UAV}},R_{i,m}^{\mathrm{HAPS}}$ & access / backhaul rate & \eqref{eq:a2g_rate}, \eqref{eq:haps_rate} & $f_i^{\mathrm{loc}},f_i^{\mathrm{UAV}},f_i^{\mathrm{HAPS}}$ & allocated CPU frequencies & \eqref{eq:opt_iot_cpu}--\eqref{eq:opt_haps_cpu} \\
$F_{\max}^{(\cdot)}$ & max.\ aggregate compute rate per tier & Sec.~\ref{sec:comp}, \eqref{eq:opt_iot_cpu} & $\kappa$ & CMOS switched-capacitance coeff. & \eqref{eq:compute_energy} \\
\midrule
\multicolumn{6}{@{}l}{Delay and energy}\\
$T_i^{\mathrm{total}}$ & task makespan (total latency) & \eqref{eq:task_completion} & $E_{i,\mathrm{tx}}^{\ell},E_{i,\mathrm{comp}}^{\ell}$ & per-phase tx / compute energy & \eqref{eq:tx_energy}, \eqref{eq:compute_energy} \\
$E_{\mathrm{sys}}$ & total IoT$+$UAV energy & \eqref{eq:system_energy} & $E_t,\overline{E}$ & per-step energy, energy budget & \eqref{eq:avg_energy} \\
$E_{\max}^{\mathrm{IoT}},E_{\max}^{\mathrm{UAV}}$ & battery capacity & \eqref{eq:opt_iot_energy}, \eqref{eq:opt_uav_energy} &  &  &  \\
\midrule
\multicolumn{6}{@{}l}{Decision variables and objective}\\
$\mathcal{A},\mathcal{W},\mathcal{P},\mathcal{F}$ & split / bandwidth / power / CPU decision sets & \eqref{eq:opt_obj} & $\mathbb{I}_i^{\mathrm{succ}}$ & deadline-success indicator & \eqref{eq:opt_obj} \\
\midrule
\bottomrule
\end{tabular}}
\end{table*}

\section{System Model and Problem Formulation}
\label{sec:system_model}
\begin{figure}
\centering
\includegraphics[width=\columnwidth]{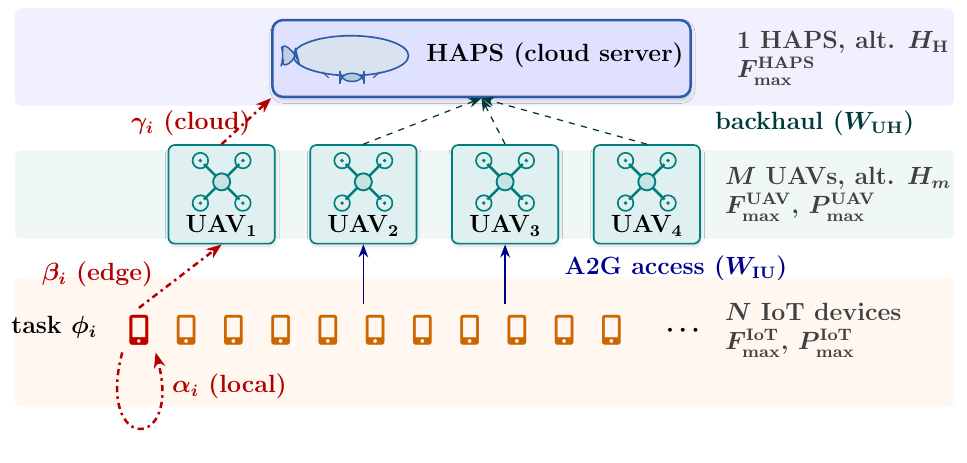}
\caption{\small{Three-tier SAGIN architecture: each task is split ($\alpha_i{+}\beta_i{+}\gamma_i{=}1$) into a local share, a UAV share sent over the $W_{\mathrm{IU}}$ access link, and a HAPS share relayed over  $W_{\mathrm{UH}}$ backhaul.}}
\label{fig:system_arch}
\end{figure}

We consider the three-tier HAPS-assisted edge computing system in Fig.~\ref{fig:system_arch}: $N$ IoT devices $\mathcal{N}=\{1,\ldots,N\}$ on the ground, $M$ UAVs $\mathcal{M}=\{1,\ldots,M\}$ as aerial edge servers, and one HAPS as the cloud tier. Each task is served by one UAV, which receives its offloaded data, executes its edge portion, and relays its cloud portion to the HAPS, so a task may be divided among the three tiers. Table~\ref{tab:notation} summarizes the notation and gives the equation or section where each symbol is defined.

\subsection{Deployment and Device Association} \label{sec:association}
\begin{figure}[!t]
\centering
\includegraphics[width=0.8\columnwidth]{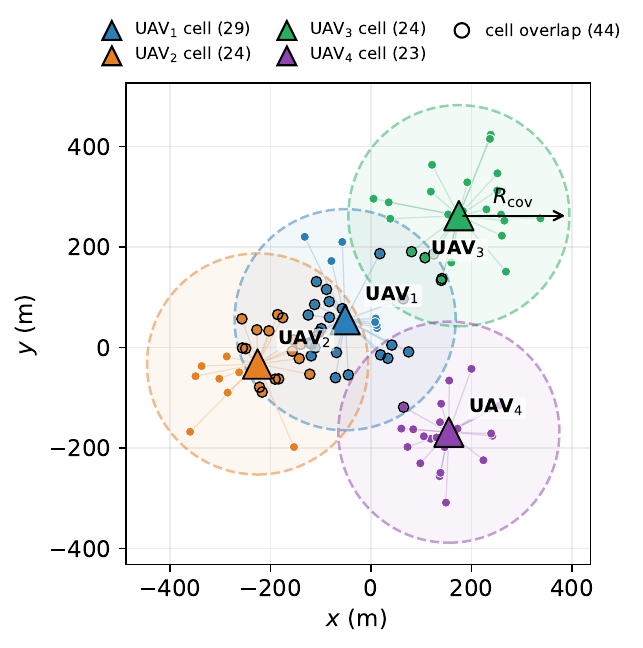}
\caption{\small{Representative Gaussian-hotspot deployment with $k$-means-positioned UAVs and coverage-based association.}}
\label{fig:deployment}
\end{figure}

To capture the spatial clustering commonly observed in IoT deployments, the device locations are drawn from a mixture of $K_{\mathrm{h}}$ Gaussian hotspots over the rectangular service area $\mathcal{R}$. The horizontal position $\mathbf{x}_n$ of device $n$ follows
\begin{equation}
\mathbf{x}_n
\sim
\sum_{q=1}^{K_{\mathrm{h}}}
\pi_q\,
\mathcal{N}\!\left(
\boldsymbol{\mu}_q,\mathbf{\Sigma}_q
\right),
\label{eq:iot_deployment}
\end{equation}
where $\boldsymbol{\mu}_q$, $\mathbf{\Sigma}_q$, and $\pi_q$ denote the center, spatial spread, and sampling probability of hotspot $q$, respectively. The UAVs are positioned according to the resulting device distribution rather than at predefined locations: $k$-means clustering is applied to the IoT coordinates and one UAV is placed above each cluster centroid at altitude $H_m$, which puts the aerial edge nodes near the regions of highest device density. The HAPS is located above the center of $\mathcal{R}$ at altitude $H_{\mathrm{H}}$ and serves all UAVs as the common cloud tier. Fig.~\ref{fig:deployment} shows a representative deployment with the device distribution, the UAV locations, and the coverage regions of radius $R_{\mathrm{cov}}$ that define the feasible UAV set of each task.
Each UAV covers a horizontal radius $R_{\mathrm{cov}}$, so the feasible UAV set of task $i$ is
\begin{equation}
\mathcal{M}_i=
\left\{
m\in\mathcal{M}:
d_{i,m}^{\mathrm{2D}}\leq R_{\mathrm{cov}}
\right\},
\label{eq:feasible_uav}
\end{equation}
where $d_{i,m}^{\mathrm{2D}}$ is the horizontal distance between the source device of task $i$ and UAV $m$. When several UAVs are feasible for a task, the serving UAV is selected by the auction mechanism adopted from our prior work~\cite{cai2026prior}. The auction is performed once per macro-slot for the batch of newly arrived tasks. For task $i$, the urgency score and the cost of assigning it to UAV $m$ are:
\begin{equation}
u_i={D_i}/{T_i^{\max}},
\qquad
c_{i,m}=D_i\left(1+{d_{i,m}^{\mathrm{2D}}}/{d_c}\right),
\label{eq:auction_scores}
\end{equation}
where $D_i$ is the task input size, $T_i^{\max}$ its latency deadline, $d_{i,m}^{\mathrm{2D}}$ the horizontal distance between the source device and UAV $m$, and $d_c$ a distance scale. Larger tasks with tighter deadlines therefore receive higher urgency, while the association cost grows with both task size and distance.

The urgency and cost values are independently min-max normalized within each arrival batch, giving $\widetilde{u}_i$ and $\widetilde{c}_{i,m}$. A feasible UAV bids for task $i$ according to the normalized net utility $\widetilde{u}_i-\widetilde{c}_{i,m}$ after subtracting the task's current auction price. To balance the load, UAVs that have not reached a task limit for the batch are considered first; if all feasible UAVs have reached the limit, the restriction is lifted so that a task within coverage is never rejected. The UAV with the highest bid wins, and the task price is increased by the winning margin over the second-highest bid plus a small increment, adapted from the classical auction rule~\cite{bertsekas1988auction}. The selected UAV remains fixed for the lifetime of the task. The auction determines only the serving UAV; the split among local, UAV, and HAPS execution, together with the comm. and computation resource allocation, is controlled by the proposed learning framework.

\subsection{Task Arrival and Queueing Model}
\label{sec:traffic}
The system operates on two timescales: tasks arrive and are split every macro-slot of duration $\Delta T$, while comm. and computation resources are reallocated every micro-slot of duration $\Delta t$, with $\Delta T=K_{\mathrm{slots}}\Delta t$ and the global micro-slot index $\tau=(t-1)K_{\mathrm{slots}}+k$, $k\in\{1,\ldots,K_{\mathrm{slots}}\}$. Arrivals follow a Poisson process with aggregate rate $\Lambda$, and each task is characterized by
\begin{equation}
\phi_i
\triangleq
\left(
D_i,C_i,T_i^{\max},\tau_i^{\mathrm{arr}}
\right),
\label{eq:task_model}
\end{equation}
its input size, CPU cycles per input bit, latency deadline, and arrival micro-slot. Tasks that arrive within a macro-slot are admitted at the next macro-slot boundary, when their deadline clocks start. Source devices are sampled without replacement within each macro-slot, so a device generates at most one new task per macro-slot, while tasks admitted in earlier macro-slots may remain active concurrently in their queues until they complete or their deadlines expire. Each task is then divided among local, UAV, and HAPS execution according to
\begin{equation}
\alpha_i+\beta_i+\gamma_i=1,
\qquad
\alpha_i,\beta_i,\gamma_i\in[0,1],
\label{eq:task_split}
\end{equation}
where fractions below the minimum schedulable threshold are set to zero and the remaining fractions are re-normalized so that they still sum to one. The resulting split remains fixed throughout the task lifetime, whereas comm. and computation resources are reallocated every micro-slot. Work that is not finished within a micro-slot stays in the queue of its resource. For resource $j$, $Q_j[\tau]$ is the number of unfinished phases waiting for or using the resource, and $B_j[\tau]$ is their total remaining workload. A completed phase leaves the queue and, if a further phase is required, joins the queue of the next resource. Once admitted, a task stays in the system until all of its active branches complete or its deadline expires; a task is successful only if every nonzero branch finishes within $T_i^{\max}$, otherwise it is dropped at its deadline and all resources assigned to it are released. The success rate is measured over all generated tasks: a task can be lost either through an admission drop (which only the flat baselines incur) or through a deadline miss, and each lost task counts as one failed task.

\subsubsection{Service Model}
Because bandwidth, power, CPU, and interference change from one micro-slot to the next, service is tracked through accumulated bits or cycles rather than through a delay computed at a constant rate. With the phases
\begin{equation}
\ell\in
\{\mathrm{a2g},\mathrm{bh},\mathrm{loc},
\mathrm{uav},\mathrm{haps}\}
\label{eq:phase}
\end{equation}
(air-to-ground (A2G) transmission, UAV-HAPS backhaul, local, UAV, and HAPS computing), the cumulative service received by phase $\ell$ of task $i$ is
\begin{equation}
S_i^\ell[\tau]
=
\sum_{\tau'=\tau_i^{\ell,\mathrm{start}}}^{\tau}
r_i^\ell[\tau']\Delta t,
\label{eq:cumulative_service}
\end{equation}
where $r_i^\ell[\tau]$ is the instantaneous transmission rate of a comm. phase or the allocated CPU frequency of a computation phase, and the phase completes at
\begin{equation}
\tau_i^{\ell,\mathrm{fin}}
=
\min
\left\{
\tau:
S_i^\ell[\tau]\geq W_i^\ell
\right\},
\label{eq:phase_completion}
\end{equation}
with the phase workloads
\begin{align}
W_i^{\mathrm{a2g}} = (\beta_i+\gamma_i)D_i, \;\;
W_i^{\mathrm{bh}} = \gamma_iD_i, \nonumber\\
W_i^{\mathrm{loc}} = \alpha_iD_iC_i, \;\;
W_i^{\mathrm{uav}} = \beta_iD_iC_i,\;\;
W_i^{\mathrm{haps}} = \gamma_iD_iC_i.
\label{eq:phase_workloads}
\end{align}
Local execution and A2G transmission start at admission, UAV computing and the backhaul start after the A2G phase, and HAPS computing after the backhaul (Fig.~\ref{fig:system_arch}):
\begin{equation}
\tau_i^{\ell,\mathrm{start}}
=
\begin{cases}
\tau_i^{\mathrm{arr}},
& \ell\in\{\mathrm{loc},\mathrm{a2g}\},\\
\tau_i^{\mathrm{a2g},\mathrm{fin}}+1,
& \ell\in\{\mathrm{uav},\mathrm{bh}\},\\
\tau_i^{\mathrm{bh},\mathrm{fin}}+1,
& \ell=\mathrm{haps}.
\end{cases}
\label{eq:phase_dependency}
\end{equation}
The three branches proceed in parallel once their dependencies are met, and the task latency is set by the last active branch,
\begin{equation}
T_i^{\mathrm{total}}
=
\left(
\max_{\ell\in
\{\mathrm{loc},\mathrm{uav},\mathrm{haps}\}}
\tau_i^{\ell,\mathrm{fin}}
-
\tau_i^{\mathrm{arr}}
\right)\Delta t ,
\label{eq:task_completion}
\end{equation}
where a branch with zero workload counts as completed immediately. The task succeeds if $T_i^{\mathrm{total}}\leq T_i^{\max}$.

\subsubsection{Computation Model}
\label{sec:comp}
Let $f_i^{\mathrm{loc}}[t,k]$, $f_i^{\mathrm{UAV}}[t,k]$, and $f_i^{\mathrm{HAPS}}[t,k]$ denote the CPU frequencies allocated to task $i$ at the three tiers; they are the service rates $r_i^{\mathrm{loc}}$, $r_i^{\mathrm{uav}}$, and $r_i^{\mathrm{haps}}$ of the computation phases. In every micro-slot, the total allocation at a device, a UAV, or the HAPS is bounded by its capacity $F_{\max}^{\mathrm{IoT}}$, $F_{\max}^{\mathrm{UAV}}$, or $F_{\max}^{\mathrm{HAPS}}$.

\subsection{Channel Model}
\label{sec:channel}
As in \cite{cai2026prior}, we assume the IoT-UAV access link is obstructed by buildings, whereas the UAV-HAPS backhaul is an almost unobstructed air-to-air (A2A) link, so the two are modeled separately.  For the access link we use the probabilistic A2G model~\cite{alhourani2014los}, in which the line-of-sight (LoS) probability depends on the elevation angle $\theta_{i,m}=\frac{180}{\pi}\arcsin(H_m/d_{i,m})$ (in degrees) between device $i$ and UAV $m$ at 3D distance $d_{i,m}$:
\begin{equation}
P_{\mathrm{LoS}}(\theta_{i,m})
=
\frac{1}
{1+a\exp[-b(\theta_{i,m}-a)]},
\label{eq:plos}
\end{equation}
where $a$ and $b$ depend on the environment and $P_{\mathrm{NLoS}}=1-P_{\mathrm{LoS}}$. The average path loss~\cite{alhourani2014pathloss} and the resulting channel gain are
\begin{equation}
\overline{L}_{i,m}
=
P_{\mathrm{LoS}}
\left(
L_{\mathrm{FS}}+\eta_{\mathrm{LoS}}
\right)
+
P_{\mathrm{NLoS}}
\left(
L_{\mathrm{FS}}+\eta_{\mathrm{NLoS}}
\right),
\label{eq:pathloss}
\end{equation}
\begin{equation}
h_{i,m}
=
10^{-\overline{L}_{i,m}/10},
\label{eq:a2g_gain}
\end{equation}
where $L_{\mathrm{FS}}=20\log_{10}({4\pi f_c d_{i,m}}/{c})$ is the free-space path loss at carrier frequency $f_c$, $c$ is the speed of light, and $\eta_{\mathrm{LoS}}$ and $\eta_{\mathrm{NLoS}}$ are the excess attenuation of the two propagation conditions. The gain is fixed by the deployment geometry within an episode, but the access rate still varies from one micro-slot to the next through power, bandwidth, and interference. The UAV-HAPS backhaul is an A2A link dominated by the LoS path, with gain
\begin{equation}\small
h_{m,\mathrm{H}}
=
G_{\mathrm{A2A}}\rho_0
(
{d_{m,\mathrm{H}}}/{d_0}
)^{-\alpha_{\mathrm{AA}}},
\label{eq:a2a_gain}
\end{equation}
where $\rho_0=(c/4\pi f_c)^2$ is the reference gain at $d_0=1$~m, $\alpha_{\mathrm{AA}}$ is the A2A path loss exponent, and $G_{\mathrm{A2A}}$ is the combined gain of the directional antennas, which compensates the large free-space attenuation of the link. On the backhaul, the HAPS altitude is much larger than the horizontal separation between a UAV and the HAPS ground projection, so the UAV-HAPS distance $d_{m,\mathrm{H}}$ is dominated by the vertical distance and the backhaul channel gains are nearly identical across UAVs for a fixed deployment.

\subsection{Communication Model}
\label{sec:comm}
Following the model in \cite{cai2026prior}, the access and backhaul links use separate bands, so they do not interfere with each other, and bandwidth and transmit power are reallocated every micro-slot. For task $i$ served by UAV $m$ at micro-slot $[t,k]$, let $w_i^{\mathrm{U}}[t,k]\in[0,1]$ be its fraction of the access bandwidth $W_{\mathrm{IU}}$ and $p_i^{\mathrm{IoT}}[t,k]$ the transmit power of its source device. The uplink rate~\cite{goldsmith2005wireless} is
\begin{equation}\small
R_{i,m}^{\mathrm{UAV}}[t,k]
=
w_i^{\mathrm{U}}[t,k]W_{\mathrm{IU}}
\log_2
\left(
1+\Gamma_{i,m}^{\mathrm{U}}[t,k]
\right),
\label{eq:a2g_rate}
\end{equation}
with the SINR
\begin{equation}\small
\Gamma_{i,m}^{\mathrm{U}}[t,k]
=
\frac{
p_i^{\mathrm{IoT}}[t,k]h_{i,m}
}{
N_0 w_i^{\mathrm{U}}[t,k]W_{\mathrm{IU}}
+
I_{i,m}[t,k]
},
\label{eq:a2g_sinr}
\end{equation}
where $N_0$ is the noise power spectral density. Tasks associated with the same UAV are separated through FDMA and do not interfere with one another. Across different UAV cells, however, the access spectrum is allocated independently, so sub-bands may overlap. The interference is modeled through the expected overlap between independently placed sub-bands. For an interfering task $j$ associated with UAV $v\neq m$, let $g_{j,m}$ denote the channel gain from its source device to UAV $m$, evaluated with the same air-to-ground model as the desired link but at the distance between the interfering device and UAV $m$. Its received power spectral density at UAV $m$ is
\begin{equation}
\psi_{j,m}[t,k]
=
\frac{
p_j^{\mathrm{IoT}}[t,k]g_{j,m}
}{
w_j^{\mathrm{U}}[t,k]W_{\mathrm{IU}}
}.
\label{eq:interference_psd}
\end{equation}
Under independent uniform sub-band placement, the expected bandwidth overlap between tasks $i$ and $j$ is
$\mathbb{E}[B_{i,j}^{\mathrm{overlap}}]=w_i^{\mathrm{U}}[t,k]\,w_j^{\mathrm{U}}[t,k]\,W_{\mathrm{IU}}$,
so the expected interference contributed by task $j$ is
\begin{equation}
I_{j\rightarrow i}[t,k]
=
\psi_{j,m}[t,k]\,\mathbb{E}[B_{i,j}^{\mathrm{overlap}}]
=
w_i^{\mathrm{U}}[t,k]\,
p_j^{\mathrm{IoT}}[t,k]g_{j,m}.
\label{eq:pair_interference}
\end{equation}
Summing over the active transmissions in all neighboring cells gives the expected inter-cell interference used in the access-link SINR~\eqref{eq:a2g_sinr},
\begin{equation}
I_{i,m}[t,k]
=
w_i^{\mathrm{U}}[t,k]
\sum_{v\in\mathcal{M}\setminus\{m\}}
\sum_{j\in\mathcal{N}_v[t,k]}
p_j^{\mathrm{IoT}}[t,k]g_{j,m},
\label{eq:total_interference}
\end{equation}
where $\mathcal{N}_v[t,k]$ is the set of active tasks transmitting to UAV $v$ at micro-slot $[t,k]$. This term is the expected interference seen by the controller, which allocates bandwidth fractions rather than explicit sub-carriers. Tasks forwarded to the HAPS share the backhaul bandwidth $W_{\mathrm{UH}}$ without mutual interference, with fractions $w_i^{\mathrm{H}}[t,k]$, $\sum_i w_i^{\mathrm{H}}[t,k]\leq1$, and the transmit power $p_{i,m}^{\mathrm{UAV}}[t,k]$ of the serving UAV, which gives
\begin{equation}\small
R_{i,m}^{\mathrm{HAPS}}[t,k]
=
w_i^{\mathrm{H}}[t,k]W_{\mathrm{UH}}
\log_2
\left(
1+\Gamma_{i,m}^{\mathrm{H}}[t,k]
\right),
\label{eq:haps_rate}
\end{equation}
\begin{equation}\small
\Gamma_{i,m}^{\mathrm{H}}[t,k]
=
\frac{
p_{i,m}^{\mathrm{UAV}}[t,k]h_{m,\mathrm{H}}
}{
N_0 w_i^{\mathrm{H}}[t,k]W_{\mathrm{UH}}
}.
\label{eq:haps_sinr}
\end{equation}
The rates \eqref{eq:a2g_rate} and \eqref{eq:haps_rate} are the service rates $r_i^{\mathrm{a2g}}[\tau]$ and $r_i^{\mathrm{bh}}[\tau]$ in \eqref{eq:cumulative_service}.

\subsection{Energy Consumption Model}
\label{sec:energy}
We account for the comm. and computation energy of the battery-powered IoT devices and UAVs, accumulated only while a phase is active. With the dynamic voltage and frequency scaling model~\cite{burd1996dvfs, wangJointOffloadingComputing2018}, in which the CPU power grows with $f^3$, the computation and transmission energy of phase $\ell$ of task $i$ are
\begin{equation}
E_{i,\mathrm{comp}}^{\ell}
=
\sum_{\tau\in\mathcal{T}_i^{\ell}}
\kappa
\left(f_i^{\ell}[\tau]\right)^3
\Delta t,
\qquad
\ell\in\{\mathrm{loc},\mathrm{uav},\mathrm{haps}\},
\label{eq:compute_energy}
\end{equation}
\begin{equation}
E_{i,\mathrm{tx}}^{\ell}
=
\sum_{\tau\in\mathcal{T}_i^{\ell}}
p_i^{\ell}[\tau]\Delta t,
\qquad
\ell\in\{\mathrm{a2g},\mathrm{bh}\},
\label{eq:tx_energy}
\end{equation}
where $\kappa$ is the effective switched capacitance, $\mathcal{T}_i^{\ell}$ is the set of micro-slots in which the phase is active, $p_i^{\mathrm{a2g}}=p_i^{\mathrm{IoT}}$, $p_i^{\mathrm{bh}}=p_{i,m}^{\mathrm{UAV}}$, and the transmit power is zero whenever no bandwidth is assigned to the link. The solar-powered HAPS~\cite{kurt2021hapsmag,lou2023hapsedge} is excluded from the energy objective, as are UAV propulsion and hovering, which do not depend on the offloading decisions; cloud offloading still costs UAV energy over the backhaul. The system energy is therefore
\begin{equation}
E_{\mathrm{sys}}
=
\sum_{i\in\mathcal{N}_{\mathrm{all}}}
\left(
E_{i,\mathrm{comp}}^{\mathrm{loc}}
+
E_{i,\mathrm{tx}}^{\mathrm{a2g}}
+
E_{i,\mathrm{comp}}^{\mathrm{uav}}
+
E_{i,\mathrm{tx}}^{\mathrm{bh}}
\right).
\label{eq:system_energy}
\end{equation}

\subsection{Problem Formulation}
\label{sec:problem_formulation}
We jointly optimize the task split and the allocation of comm. and computation resources to maximize the fraction of generated tasks completed within their deadlines. Let $\mathcal{N}_{\mathrm{all}}$ denote the set of tasks generated over the considered horizon. The decision variables are grouped as $\mathcal{A}=\{\alpha_i,\beta_i,\gamma_i\}$,
$\mathcal{W}=\{w_i^{\mathrm{U}}[t,k],w_i^{\mathrm{H}}[t,k]\}$, 
$\mathcal{P}=\{p_i^{\mathrm{IoT}}[t,k],p_{i,m}^{\mathrm{UAV}}[t,k]\}$, 
$\mathcal{F}=\{f_i^{\mathrm{loc}}[t,k],f_i^{\mathrm{UAV}}[t,k], f_i^{\mathrm{HAPS}}[t,k]\}$.
At each micro-slot $[t,k]$, the active sets contain only task phases that have not completed and whose deadlines have not expired. We denote by $\mathcal{N}_{m}^{\mathrm{tx}}[t,k]$ the tasks transmitting to UAV $m$, by $\mathcal{N}_{\mathrm{H}}^{\mathrm{tx}}[t,k]$ the tasks using the shared UAV-HAPS backhaul, and by $\mathcal{N}_{n}^{\mathrm{comp}}[t,k]$, $\mathcal{N}_{m}^{\mathrm{comp}}[t,k]$, and $\mathcal{N}_{\mathrm{H}}^{\mathrm{comp}}[t,k]$ the tasks computing at IoT device $n$, UAV $m$, and the HAPS, respectively. A task is removed from the corresponding sets when it completes or its deadline expires. Let $\mathbb{I}_i^{\mathrm{succ}}=1$ if all active branches of task $i$ complete within $T_i^{\max}$, and $\mathbb{I}_i^{\mathrm{succ}}=0$ otherwise. The joint optimization problem is formulated as:
\begin{subequations}
\label{eq:opt_prob}
\begin{align} \small
\max_{\mathcal{A},\mathcal{W},\mathcal{P},\mathcal{F}}
\quad &\small
\mathbb{E}\!\left[
\frac{1}{|\mathcal{N}_{\mathrm{all}}|}
\sum_{i\in\mathcal{N}_{\mathrm{all}}}
\mathbb{I}_i^{\mathrm{succ}}
\right]
\label{eq:opt_obj}
\\
\mathrm{s.t.}\quad
&\small
\alpha_i+\beta_i+\gamma_i=1,
\;\forall i\in\mathcal{N}_{\mathrm{all}},
\label{eq:opt_split}
\\
&\small
\alpha_i,\beta_i,\gamma_i\in[0,1],
\;\forall i\in\mathcal{N}_{\mathrm{all}},
\label{eq:opt_split_range}
\\
&\small
\sum_{i\in\mathcal{N}_{m}^{\mathrm{tx}}[t,k]}
w_i^{\mathrm{U}}[t,k]\leq1,
\;\forall m,t,k,
\label{eq:opt_access_bw}
\\
&\small
\sum_{i\in\mathcal{N}_{\mathrm{H}}^{\mathrm{tx}}[t,k]}
w_i^{\mathrm{H}}[t,k]\leq1,
\;\forall t,k,
\label{eq:opt_backhaul_bw}
\\
&\small
\sum_{i\in\mathcal{N}_{n}^{\mathrm{tx}}[t,k]}
p_i^{\mathrm{IoT}}[t,k]
\leq P_{\max}^{\mathrm{IoT}},
\;\forall n,t,k,
\label{eq:opt_iot_power}
\\
&\small
\sum_{i\in\mathcal{N}_{m,\mathrm{H}}^{\mathrm{tx}}[t,k]}
p_{i,m}^{\mathrm{UAV}}[t,k]
\leq P_{\max}^{\mathrm{UAV}},
\;\forall m,t,k,
\label{eq:opt_uav_power}
\\
&\small
\sum_{i\in\mathcal{N}_{n}^{\mathrm{comp}}[t,k]}
f_i^{\mathrm{loc}}[t,k]
\leq F_{\max}^{\mathrm{IoT}},
\;\forall n,t,k,
\label{eq:opt_iot_cpu}
\\
&\small
\sum_{i\in\mathcal{N}_{m}^{\mathrm{comp}}[t,k]}
f_i^{\mathrm{UAV}}[t,k]
\leq F_{\max}^{\mathrm{UAV}},
\;\forall m,t,k,
\label{eq:opt_uav_cpu}
\\
&\small
\sum_{i\in\mathcal{N}_{\mathrm{H}}^{\mathrm{comp}}[t,k]}
f_i^{\mathrm{HAPS}}[t,k]
\leq F_{\max}^{\mathrm{HAPS}},
\;\forall t,k,
\label{eq:opt_haps_cpu}
\\
&\small
\sum_{i\in\Omega_n^{\mathrm{IoT}}}
\left(
E_{i,\mathrm{comp}}^{\mathrm{loc}}
+
E_{i,\mathrm{tx}}^{\mathrm{a2g}}
\right)
\leq E_{\max}^{\mathrm{IoT}},
\;\forall n,
\label{eq:opt_iot_energy}
\\
&\small
\sum_{i\in\Omega_m^{\mathrm{UAV}}}
\left(
E_{i,\mathrm{comp}}^{\mathrm{uav}}
+
E_{i,\mathrm{tx}}^{\mathrm{bh}}
\right)
\leq E_{\max}^{\mathrm{UAV}},
\;\forall m,
\label{eq:opt_uav_energy}
\\
&\small
w_i^{\mathrm{U}},w_i^{\mathrm{H}},
p_i^{\mathrm{IoT}},p_{i,m}^{\mathrm{UAV}},
f_i^{\mathrm{loc}},f_i^{\mathrm{UAV}},
f_i^{\mathrm{HAPS}}\geq0.
\label{eq:opt_nonnegative}
\end{align}
\end{subequations}

Here, $\Omega_n^{\mathrm{IoT}}$ contains the tasks generated by device $n$ and $\Omega_m^{\mathrm{UAV}}$ the tasks associated with UAV $m$. Constraints \eqref{eq:opt_split}-\eqref{eq:opt_split_range} keep the whole workload of each task across the local, UAV, and HAPS branches. Constraints \eqref{eq:opt_access_bw}-\eqref{eq:opt_backhaul_bw} bound the access and backhaul bandwidth, \eqref{eq:opt_iot_power}-\eqref{eq:opt_uav_power} the instantaneous transmit power of the IoT devices and UAVs, and \eqref{eq:opt_iot_cpu}-\eqref{eq:opt_haps_cpu} the total CPU allocation at each computing node. Finally, \eqref{eq:opt_iot_energy}-\eqref{eq:opt_uav_energy} bound the battery energy that the IoT devices and UAVs may spend.

The power and energy constraints play different roles: the power constraints limit what a node may transmit at any instant, whereas the energy constraints limit what its battery may spend over the whole horizon. An online controller, however, does not know the future task arrivals, queue states, or energy demands when it makes the current decision, so the battery constraints cannot be enforced directly. We instead use an average energy constraint as a tractable online approximation. Let $E_t$ denote the total comm. and computation energy consumed by the IoT and UAV tiers during macro-slot $t$. The online controller is subject to
\begin{equation}\small
\frac{1}{T}
\sum_{t=1}^{T}
\mathbb{E}[E_t]
\leq \overline{E},
\label{eq:avg_energy}
\end{equation}
where $\overline{E}$ is the allowed average energy consumption per macro-slot. This constraint does not reproduce the battery limits exactly, but it can be enforced online, it controls the energy spent over the long run, and it enters the constrained learning framework of the next section.

Problem~\eqref{eq:opt_prob} is a non-convex mixed-integer nonlinear program (MINLP). The discrete part comes from the success indicator $\mathbb{I}_i^{\mathrm{succ}}$ and the slotted completion times, while the continuous split and allocation variables are coupled nonlinearly through the SINR and rate expressions. The underlying offloading problem is combinatorial and NP-hard even under simplified resource settings~\cite{huang2019deepoffload, chen2016multiuser}. Moreover, random arrivals, completions, and deadline expiration keep changing the active task and resource sets, so a static optimization is unsuitable for online control. These challenges motivate the hierarchical graph DRL framework of Sec.~\ref{sec:framework}.

\section{Proposed Framework: AeroGLIDER}
\label{sec:framework}
To solve the stochastic two-timescale problem of Sec.~\ref{sec:problem_formulation}, we propose AeroGLIDER, a hierarchical DRL framework that acts on graphs. ``Aero'' refers to the aerial computing tiers formed by the UAVs and the HAPS.

\subsection{Framework Overview}
\label{sec:overview}
\begin{figure*}[!t]
\centering
\includegraphics[width=0.7\textwidth]{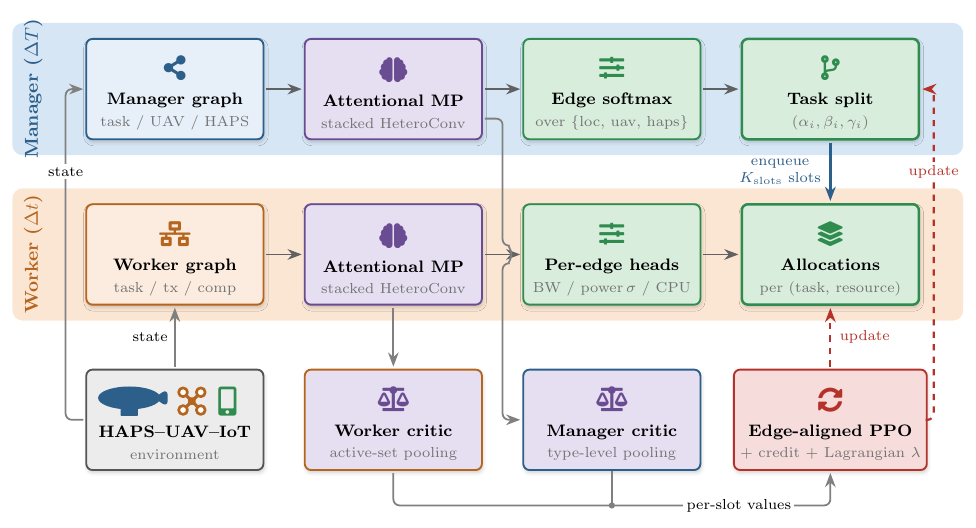}
\caption{AeroGLIDER overview: two heterogeneous graphs, message passing with attention, actor heads on the graph edges, and a relational critic whose advantage is shared by the edge actions in the PPO update.}
\label{fig:framework}
\end{figure*}

\begin{figure*}[!t]
\centering
\includegraphics[width=0.7\textwidth]{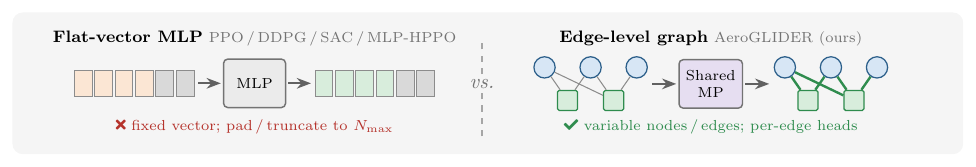}
\caption{Flat vector versus graph: left, the flat baselines map a state vector of fixed length to an action vector of fixed width; right, the AeroGLIDER graph grows with the instance.}
\label{fig:interface}
\end{figure*}

AeroGLIDER models the online control problem as a two-timescale constrained Markov decision process (CMDP), summarized in Fig.~\ref{fig:framework}. The manager acts once per macro-slot and sets the split $(\alpha_i,\beta_i,\gamma_i)$ of each newly admitted task. The worker acts once per micro-slot and allocates the comm. and computation resources that the active task phases need. Each decision is taken on a heterogeneous graph. The manager graph $\mathcal{G}_t^{\mathrm{m}}$ connects the newly admitted tasks to the aerial resources that matter for their split. The worker graph $\mathcal{G}_{\tau}^{\mathrm{w}}$ connects the active transmission and computation phases to the resources they compete for, and carries the residual workload, deadline slack, queue depth, and backlog. Two heterogeneous message-passing layers with attention encode these graphs. A head per task produces the three-way split, and a head per edge produces the bandwidth, transmit power, and CPU allocations of the active task-resource pairs. Fig.~\ref{fig:interface} contrasts this graph interface with the flat baselines, whose state and action vectors have a fixed size. Each agent has its own critic on the graph, the energy constraint enters the manager reward through Lagrangian relaxation, and both policies are trained with hierarchical PPO. The following subsections detail each component.

\subsection{Two-Timescale Constrained MDP}
\label{sec:cmdp}
We formalize AeroGLIDER as two coupled decision processes operating at different timescales. Let ${\mathcal{G}}_t^{\mathrm{m}}\in\mathcal{S}^{\mathrm{m}}$ and
$\mathcal{G}_\tau^{\mathrm{w}}\in\mathcal{S}^{\mathrm{w}}$
denote the manager and worker states, respectively. The manager policy operates once per macro-slot as
\begin{equation}
\pi_{\theta_{\mathrm{m}}}^{\mathrm{m}}
\left(
\mathbf{a}_t^{\mathrm{m}}
\mid
\mathcal{G}_t^{\mathrm{m}}
\right),
\end{equation}
where $\mathbf{a}_t^{\mathrm{m}}$ contains the partitions
$(\alpha_i,\beta_i,\gamma_i)$ of all tasks admitted at macro-slot
$t$. The worker policy operates at every micro-slot as
\begin{equation}
\pi_{\theta_{\mathrm{w}}}^{\mathrm{w}}
\left(
\mathbf{a}_\tau^{\mathrm{w}}
\mid
\mathcal{G}_\tau^{\mathrm{w}}
\right),
\end{equation}
where $\mathbf{a}_\tau^{\mathrm{w}}$ contains the bandwidth, transmit power, and CPU decisions for the active task-resource pairs.

A manager action fixes the splits whose workloads are then served by the next $K_{\mathrm{slots}}$ worker decisions, and each worker action updates the residual workloads, queue states, energy, and task status that form the next states. Because the graph states carry everything about past actions that matters for the next decision, both policies are memoryless and conditioned only on the current graph.
\subsubsection{Task and Energy Reward}
Let $S_t$ denote the number of tasks completed within their deadlines during macro-slot $t$, $F_t$ the number of tasks dropped upon deadline expiration, and $A_t$ the number of active tasks. The task reward is:
\begin{equation}\small
r_t^{\mathrm{task}}= \frac{2S_t-F_t}{\max(1,A_t)},
\label{eq:task_reward}
\end{equation}
which encourages timely task completion while penalizing deadline
violations.

The average energy constraint of Sec.~\ref{sec:problem_formulation} is handled by Lagrangian relaxation, as in constrained RL \cite{paternain2019safe}. The manager reward is
\begin{equation}\small
r_t^{\mathrm{m}}= r_t^{\mathrm{task}}-\lambda_t\left(E_t-\overline{E}
\right),
\label{eq:manager_reward}
\end{equation}
where $E_t$ is the total IoT and UAV energy consumed during macro-slot $t$, $\overline{E}$ is the energy budget, and $\lambda_t\geq0$ is the Lagrange multiplier. The multiplier is updated by projected dual ascent,
\begin{equation}
\lambda_{t+1}
=
\left[
\lambda_t+
\eta_{\lambda}
\left(
E_t-\overline{E}
\right)
\right]_{0}^{\lambda_{\max}},
\label{eq:lambda_update}
\end{equation}
where $\eta_{\lambda}$ is the dual step size and $[\cdot]_{0}^{\lambda_{\max}}$ denotes projection onto $[0,\lambda_{\max}]$. Over an episode of $T$ macro-slots, the coupled policies seek to maximize the expected discounted return
\begin{equation}
J
=
\mathbb{E}_{\pi^{\mathrm{m}},\pi^{\mathrm{w}}}
\left[
\sum_{t=1}^{T}
\gamma_{\mathrm{m}}^{\,t-1}
r_t^{\mathrm{m}}
\right],
\label{eq:joint_return}
\end{equation}
where $\gamma_{\mathrm{m}}$ is the macro-slot discount factor.

\subsection{Heterogeneous Graph Construction}
\label{sec:graphs}
Both graphs are rebuilt as tasks arrive, progress, complete, or expire. Since the node features have different physical units, each continuous feature $z$ is scaled as
\begin{equation}
\bar z={z}/{\rho_z},
\label{eq:feature_scaling}
\end{equation}
where $\rho_z>0$ is a fixed reference scale (a system constant, not a statistic of the training data) that keeps the continuous inputs of order one; bounded ratios and binary indicators are used as they are. The reference scales are listed in Table~\ref{tab:hype}.

\subsubsection{Manager Graph}
At macro-slot $t$, let $\mathcal{B}(t)$ denote the set of newly admitted tasks. The manager graph is defined as
$\mathcal{G}_t^{\mathrm{m}}
=(\mathcal{V}_t^{\mathrm{m}},\mathcal{E}_t^{\mathrm{m}})$, with node set
\begin{equation}
\mathcal{V}_t^{\mathrm{m}}
=
\{\mathrm{task}_i\}_{i\in\mathcal{B}(t)}
\cup
\{\mathrm{uav}_m\}_{m=1}^{M}
\cup
\{\mathrm{haps}\}.
\label{eq:manager_nodes}
\end{equation}
Each task node is connected to its serving UAV and to the HAPS, so that its embedding can take in the state of the aerial resources that matter for its split. Local execution needs no separate resource node because its feasibility is part of the task features. The feature vector of task $i$ is
\begin{equation}
\mathbf{x}_i^{\mathrm{task}}
=
\left[
\bar b_i,\;
\bar d_i,\;
\bar u_i,\;
c_i
\right],
\label{eq:manager_task_features}
\end{equation}
where $\bar b_i$, $\bar d_i$, and $\bar u_i$ are the scaled task size, horizontal IoT-UAV distance, and urgency, respectively, according to \eqref{eq:feature_scaling}. The local feasibility feature is
\begin{equation}
c_i
=
\min
\left(
1,
\frac{T_i^{\max}F_{\max}^{\mathrm{IoT}}}
{D_iC_i}
\right),
\label{eq:local_feasibility}
\end{equation}
the largest fraction of task $i$ that its local processor could complete within the deadline at full CPU rate.

The UAV and HAPS features summarize their current comm. and computation backlogs:
\begin{align}
\mathbf{x}_m^{\mathrm{uav}}
&=
\left[
\bar B_m^{\mathrm{tx}},\;
\bar B_m^{\mathrm{comp}}
\right],
\label{eq:manager_uav_features}
\\
\mathbf{x}^{\mathrm{haps}}
&=
\left[
\bar B_{\mathrm{H}}^{\mathrm{tx}},\;
\bar B_{\mathrm{H}}^{\mathrm{comp}}
\right].
\label{eq:manager_haps_features}
\end{align}
\subsubsection{Worker Graph}
At micro-slot $\tau$, the worker constructs a heterogeneous graph
$\mathcal{G}_{\tau}^{\mathrm{w}}
=(\mathcal{V}_{\tau}^{\mathrm{w}},
\mathcal{E}_{\tau}^{\mathrm{w}})$ of the current allocation state. Its node set is
\begin{equation}
\mathcal{V}_{\tau}^{\mathrm{w}}
=
\mathcal{V}_{\tau}^{\mathrm{task}}
\cup
\{\mathrm{tx}_j\}_{j=1}^{M+1}
\cup
\{\mathrm{comp}_j\}_{j=1}^{N+M+1}.
\label{eq:worker_nodes}
\end{equation}
Here, $\mathcal{V}_{\tau}^{\mathrm{task}}$ contains the pending tasks. The $M+1$ transmission nodes represent the UAV access resources and the HAPS backhaul resource, while the $N+M+1$ computation nodes represent the $N$ IoT processors, $M$ UAV edge servers, and the HAPS.

Each pending task is connected to the resource required by its current operation. A task awaiting transmission is connected to the corresponding transmission resource through a \emph{wait-tx} relation, whereas a task awaiting computation is connected to its processor through a \emph{wait-comp} relation. The serving UAV is fixed by the association of Sec.~\ref{sec:association}, so the tasks connected to a resource are exactly those contending for it, and these task-resource edges carry the worker's allocation decisions. The feature vector of a task node is
\begin{equation}
\mathbf{x}_{i}^{\mathrm{task}}
=
\left[
\bar L_i,\;
s_i,\;
\mathbb{I}_{i}^{\mathrm{tx}}
\right],
\label{eq:worker_task_features}
\end{equation}
where $\bar L_i$ is the scaled residual workload of the task's current operation (bits or cycles, each with its own reference scale), $\mathbb{I}_{i}^{\mathrm{tx}}\in\{0,1\}$ indicates a transmission ($1$) or computation ($0$) operation, and the slack ratio is
\begin{equation}
s_i
=
\frac{
\max\!\left(
0,\,
T_i^{\max}
-
(\tau-\tau_i^{\mathrm{arr}})\Delta t
\right)
}{
T_i^{\max}
},
\qquad
s_i\in[0,1],
\label{eq:slack_ratio}
\end{equation}
where $\tau_i^{\mathrm{arr}}$ is the admission micro-slot; $s_i$ falls to zero as the deadline approaches, and a task leaves the graph once it completes or expires.

The transmission and computation resource features are
\begin{align}
\mathbf{x}_{j}^{\mathrm{tx}}
&=
\left[
\mathbb{I}_{j}^{\mathrm{uav}},
\mathbb{I}_{j}^{\mathrm{haps}},
\bar B_{j}^{\mathrm{tx}},
\bar Q_{j}^{\mathrm{tx}}
\right],
\label{eq:worker_tx_features}
\\
\mathbf{x}_{j}^{\mathrm{comp}}
&=
\left[
\mathbb{I}_{j}^{\mathrm{iot}},
\mathbb{I}_{j}^{\mathrm{uav}},
\mathbb{I}_{j}^{\mathrm{haps}},
\bar B_{j}^{\mathrm{comp}},
\bar Q_{j}^{\mathrm{comp}}
\right].
\label{eq:worker_comp_features}
\end{align}
where the binary indicators give the tier of the resource, $\bar B_j^{(\cdot)}$ is its scaled total remaining workload (backlog), and $\bar Q_j^{(\cdot)}$ the scaled number of pending tasks competing for it (queue depth). Together they distinguish a resource serving a few large workloads from one shared by many small tasks; they are the explicit congestion signal of the worker.
\subsection{Heterogeneous Message Passing}
\label{sec:conv}
The manager and worker graphs contain different node and relation types. AeroGLIDER therefore uses two stacked heterogeneous message-passing layers, with a separate multi-head GATv2 operator for each relation type. At each layer, a node first aggregates
information from its connected neighbors and then combines this message with its current representation. For node $i$, the aggregated message at layer $\ell$ is
\begin{equation}
\mathbf{m}_{i}^{(\ell)}
=
\sum_{j\in\mathcal{N}(i)}
\alpha_{ij}^{(\ell)}
\mathbf{W}\mathbf{h}_{j}^{(\ell)},
\label{eq:neighbor_aggregation}
\end{equation}
where $\mathcal{N}(i)$ is the set of neighbors of node $i$, $\mathbf{h}_{j}^{(\ell)}$ the embedding of neighbor $j$, and $\alpha_{ij}^{(\ell)}$ its weight. Each relation type has its own transformation and attention parameters. The node representation is then updated as
\begin{equation}
\mathbf{h}_{i}^{(\ell+1)}
=
\mathrm{ReLU}
\left(
\mathrm{LN}
\left(
\mathbf{m}_{i}^{(\ell)}
+
\mathbf{h}_{i}^{(\ell)}
\right)
\right),
\label{eq:hetero_message_passing}
\end{equation}
where the residual connection preserves the previous node representation and LayerNorm stabilizes the feature scale.

Two layers suffice for a task to see the other tasks sharing its resource: the first layer lets a resource node summarize its competing tasks, and the second passes this summary back to the tasks.
\paragraph{Aggregation with Attention}
Mean aggregation, as in GraphSAGE~\cite{hamilton2017graphsage}, weights all neighbors equally even when their workloads and deadlines differ. AeroGLIDER instead uses GATv2~\cite{brody2022gatv2}, which learns an attention score for each connected pair from their current states:
\begin{equation}
e_{ij}^{(\ell)}
=
\mathbf{a}^{\top}
\mathrm{LeakyReLU}
\left(
\mathbf{W}
\left[
\mathbf{h}_{i}^{(\ell)}
\Vert
\mathbf{h}_{j}^{(\ell)}
\right]
\right),
\label{eq:gatv2_score}
\end{equation}
followed by
\begin{equation}
\alpha_{ij}^{(\ell)}
=
\frac{\exp(e_{ij}^{(\ell)})}
{\sum_{j'\in\mathcal{N}(i)}
\exp(e_{ij'}^{(\ell)})}.
\label{eq:gatv2_attention}
\end{equation}
Tasks competing for the same resource thus influence its representation according to their current workload and deadline state.

\subsection{Actor Heads on Tasks and Edges}
\label{sec:edge_heads}
Separate manager and worker heads, shared across tasks and edges, convert the embeddings into actions.
\subsubsection{Manager Head for the Task Split}
For each newly admitted task $i$, the manager combines the encoded representations of the task, its serving UAV $j(i)$, and the HAPS:
\begin{equation}
\boldsymbol{\ell}_{i}^{\mathrm{m}}
=
f_{\theta}^{\mathrm{m}}
\left(
\left[
\mathbf{h}_{i}^{\mathrm{task}}
\Vert
\mathbf{h}_{j(i)}^{\mathrm{uav}}
\Vert
\mathbf{h}^{\mathrm{haps}}
\right]
\right)
\in\mathbb{R}^{3},
\label{eq:manager_actor}
\end{equation}
where $f_{\theta}^{\mathrm{m}}(\cdot)$ is the shared manager head and $\Vert$ denotes concatenation, and a softmax gives the split
\begin{equation}
[\alpha_i,\beta_i,\gamma_i]
=
\operatorname{softmax}
\left(
\boldsymbol{\ell}_{i}^{\mathrm{m}}
\right),
\label{eq:manager_softmax}
\end{equation}
so that $\alpha_i+\beta_i+\gamma_i=1$.

\subsubsection{Worker Transmission Head}
At each micro-slot, the worker acts on the active edges between tasks and transmission resources. For each $(i,j)\in\mathcal{E}_{\tau}^{\mathrm{tx}}$, the embeddings of the task and of the transmission resource are combined into a bandwidth logit and a power logit:
\begin{equation}
\left[
\ell_{ij}^{\mathrm{BW}},
\ell_{ij}^{P}
\right]
=
f_{\theta}^{\mathrm{tx}}
\left(
\left[
\mathbf{h}_{i}^{\mathrm{task}}
\Vert
\mathbf{h}_{j}^{\mathrm{tx}}
\right]
\right),
\label{eq:tx_actor}
\end{equation}
where $f_{\theta}^{\mathrm{tx}}(\cdot)$ is shared across all active transmission edges. For each transmission resource $j$, a softmax over the tasks competing for it turns the bandwidth logits into shares:
\begin{equation}
a_{ij}^{\mathrm{BW}}
=
\frac{
\exp\!\left(\ell_{ij}^{\mathrm{BW}}\right)
}{
\sum_{i'\in\mathcal{N}(j)}
\exp\!\left(\ell_{i'j}^{\mathrm{BW}}\right)
},
\label{eq:bw_softmax}
\end{equation}
so the shares of each resource sum to one. Transmit power is generated differently: each transmission edge produces its own normalized power request
\begin{equation}
a_{ij}^{P}
=
\sigma\!\left(\ell_{ij}^{P}\right),
\qquad
a_{ij}^{P}\in(0,1),
\label{eq:power_action}
\end{equation}
which is scaled by the maximum power of transmitter $j$:
\begin{equation}
\tilde p_{ij}
=
P_{j}^{\max}a_{ij}^{P}.
\label{eq:requested_power}
\end{equation}
Since the requests are generated independently, their sum can exceed $P_j^{\max}$, so the environment projects the requested vector $\tilde{\mathbf{p}}_j$ onto the feasible set
\begin{equation}
\mathcal{P}_j
=
\left\{
\mathbf{p}_j:
p_{ij}\geq0,\;
\sum_{i\in\mathcal{N}(j)}p_{ij}
\leq P_j^{\max}
\right\},
\label{eq:power_feasible_set}
\end{equation}
as
\begin{equation}
\mathbf{p}_j
=
\Pi_{\mathcal{P}_j}
\left(
\tilde{\mathbf{p}}_j
\right),
\label{eq:power_projection}
\end{equation}
where $\Pi_{\mathcal{P}_j}(\cdot)$ denotes the projection.

\subsubsection{Worker Computation Head}
For each active edge $(i,k)\in\mathcal{E}_{\tau}^{\mathrm{comp}}$ between a task and a processor, the worker combines the two embeddings into a CPU logit:
\begin{equation}
\ell_{ik}^{\mathrm{CPU}}
=
f_{\theta}^{\mathrm{comp}}
\left(
\left[
\mathbf{h}_{i}^{\mathrm{task}}
\Vert
\mathbf{h}_{k}^{\mathrm{comp}}
\right]
\right),
\label{eq:cpu_actor}
\end{equation}
where $f_{\theta}^{\mathrm{comp}}(\cdot)$ is shared across all active computation edges. For each processor $k$, a softmax over the tasks competing for it turns the CPU logits into shares:
\begin{equation}
a_{ik}^{\mathrm{CPU}}
=
\frac{
\exp\!\left(\ell_{ik}^{\mathrm{CPU}}\right)
}{
\sum_{i'\in\mathcal{N}(k)}
\exp\!\left(\ell_{i'k}^{\mathrm{CPU}}\right)
},
\label{eq:cpu_softmax}
\end{equation}
so the CPU shares of each processor sum to one.

\subsection{Critics on the Graphs}
\label{sec:critic}
Each agent has its own critic, which maps its graph state to one scalar value per decision (per macro-slot for the manager, per micro-slot for the worker).
\subsubsection{Manager Critic}
Since the number of task nodes varies, the manager critic first averages the task embeddings,
\begin{equation}
\bar{\mathbf h}_{t}^{\mathrm{task}}
=
\frac{1}{|\mathcal{V}_{t}^{\mathrm{task}}|}
\sum_{i\in\mathcal{V}_{t}^{\mathrm{task}}}
\mathbf h_{i}^{\mathrm{task}},
\label{eq:manager_task_pool}
\end{equation}
and the UAV embeddings,
\begin{equation}
\bar{\mathbf h}_{t}^{\mathrm{uav}}
=
\frac{1}{M}
\sum_{m=1}^{M}
\mathbf h_{m}^{\mathrm{uav}}.
\label{eq:manager_uav_pool}
\end{equation}
and estimates the macro-slot value as
\begin{equation}
V_{\psi}^{\mathrm{m}}
\left(
\mathcal{G}_{t}^{\mathrm{m}}
\right)
=
\mathrm{MLP}_{\psi}^{\mathrm{m}}
\left(
\left[
\bar{\mathbf h}_{t}^{\mathrm{task}}
\Vert
\bar{\mathbf h}_{t}^{\mathrm{uav}}
\Vert
\mathbf h^{\mathrm{haps}}
\right]
\right).
\label{eq:manager_critic}
\end{equation}

\subsubsection{Worker Relational Critic}
The worker graph always contains all $N+2M+2$ resource nodes, most of them idle IoT processors, so a plain mean over resource nodes would be dominated by idle nodes and hide the few congested ones. The worker critic therefore pools only the active resources. For resource type $r\in\{\mathrm{tx},\mathrm{comp}\}$, let
\begin{equation}
\mathcal{A}_{r}(\tau)
=
\left\{
j:
\mathcal{N}_{\tau}(j)\neq\emptyset
\right\}
\label{eq:active_resource_set}
\end{equation}
denote the set of resources of type $r$ that are connected to at least one pending task. Their embeddings are summarized as
\begin{equation}
\bar{\mathbf h}_{\tau}^{r}
=
\frac{1}{|\mathcal{A}_{r}(\tau)|}
\sum_{j\in\mathcal{A}_{r}(\tau)}
\mathbf h_{j}^{r},
\qquad
r\in\{\mathrm{tx},\mathrm{comp}\}.
\label{eq:active_resource_pool}
\end{equation}
with a fallback to the mean over all nodes of that type if $\mathcal{A}_{r}(\tau)$ is empty. The task embeddings are pooled as
\begin{equation}
\bar{\mathbf h}_{\tau}^{\mathrm{task}}
=
\frac{1}{|\mathcal{V}_{\tau}^{\mathrm{task}}|}
\sum_{i\in\mathcal{V}_{\tau}^{\mathrm{task}}}
\mathbf h_{i}^{\mathrm{task}}.
\label{eq:worker_task_pool}
\end{equation}
The worker critic then estimates
\begin{equation}
V_{\phi}^{\mathrm{w}}
\left(
\mathcal{G}_{\tau}^{\mathrm{w}}
\right)
=
\mathrm{MLP}_{\phi}^{\mathrm{w}}
\left(
\left[
\bar{\mathbf h}_{\tau}^{\mathrm{task}}
\Vert
\bar{\mathbf h}_{\tau}^{\mathrm{tx}}
\Vert
\bar{\mathbf h}_{\tau}^{\mathrm{comp}}
\right]
\right).
\label{eq:worker_critic}
\end{equation}
The manager needs no such filter because its resource set contains only the UAVs and the HAPS; an empty task set is represented by a zero embedding so that graph batches stay aligned during training.
\subsection{Hierarchical PPO Training}
\label{sec:train}
The manager and worker are trained jointly and on-policy with proximal policy optimization (PPO), each on its own timescale (one transition per macro-slot and per micro-slot, respectively) and with its own critic.
\subsubsection{Advantage Estimation}
For each agent $x\in\{\mathrm{m},\mathrm{w}\}$, generalized advantage estimation (GAE) is computed from its rewards and critic values. The temporal difference (TD) error is
\begin{equation}
\delta_{\xi}^{x}
=
r_{\xi}^{x}
+
\gamma_x
V^{x}\!\left(\mathcal{G}_{\xi+1}^{x}\right)
-
V^{x}\!\left(\mathcal{G}_{\xi}^{x}\right),
\label{eq:td_error}
\end{equation}
and the corresponding advantage is
\begin{equation}
A_{\xi}^{x}
=
\sum_{l=0}^{L_{\xi}-1}
\left(
\gamma_x\lambda_{\mathrm{GAE}}
\right)^l
\delta_{\xi+l}^{x},
\label{eq:gae}
\end{equation}
where $\xi$ denotes the decision index at the corresponding timescale, $\gamma_x$ is the discount factor, $\lambda_{\mathrm{GAE}}$ is the GAE parameter, and $L_{\xi}$
denotes the remaining rollout length.
Since the worker produces several edge actions in one micro-slot, all actions generated from the same worker graph share the advantage of that slot:
\begin{equation}
A_{(i,j),\tau}^{\mathrm{w}}
=
A_{\tau}^{\mathrm{w}},
\qquad
(i,j)\in\mathcal{E}_{\tau}^{\mathrm{w}}.
\label{eq:edge_advantage}
\end{equation}
so the worker critic works per micro-slot, while each edge action keeps its own likelihood ratio in the PPO update.
\subsubsection{Reward Assignment from Manager to Worker}
The manager reward $r_t^{\mathrm{m}}$ arrives once per macro-slot, while the worker makes $K_{\mathrm{slots}}$ decisions in that interval. To give the worker a training signal for every micro-slot, the manager reward is distributed in proportion to the useful work completed in each micro-slot:
\begin{equation}
r_{t,\tau}^{\mathrm{w}}
=
r_t^{\mathrm{m}}
\frac{\omega_{\tau}}
{\sum_{\tau'\in\mathcal{K}_t}\omega_{\tau'}+\varepsilon},
\label{eq:worker_reward}
\end{equation}
where $\mathcal{K}_t$ denotes the micro-slots contained in macro-slot $t$, and $\varepsilon>0$ avoids division by zero. The completed work is measured as
\begin{equation}
\omega_{\tau}
=
\sum_{i\in\mathcal{B}_{\tau}}
\left(
\frac{\Delta b_i[\tau]}{10^{6}}
+
\frac{\Delta c_i[\tau]}{10^{9}}
\right),
\label{eq:useful_work}
\end{equation}
where $\Delta b_i[\tau]$ and $\Delta c_i[\tau]$ are the bits transmitted and cycles processed during micro-slot $\tau$. Equation~\eqref{eq:worker_reward} only distributes credit over time: since $r_t^{\mathrm{m}}$ already contains the success and energy terms, the worker inherits the same objective without a separate reward.

\subsubsection{PPO Updates}
For either agent, let $\pi_{\theta}$ denote the current policy and $\pi_{\theta_{\mathrm{old}}}$ the policy used to collect the trajectory. The probability ratio for an action $a$ is
\begin{equation}
r_{\theta}
=
\frac{
\pi_{\theta}(a\mid\mathcal{G})
}{
\pi_{\theta_{\mathrm{old}}}(a\mid\mathcal{G})
}.
\label{eq:ppo_ratio}
\end{equation}
The clipped PPO loss is
\begin{equation}
\mathcal{L}_{\mathrm{PPO}}
=
-
\mathbb{E}
\left[
\min
\left(
r_{\theta}A,\,
\operatorname{clip}
\left(
r_{\theta},
1-\epsilon_{\mathrm{PPO}},
1+\epsilon_{\mathrm{PPO}}
\right)A
\right)
\right],
\label{eq:ppo_loss}
\end{equation}
where $A$ is the advantage at the corresponding timescale. For the worker, the ratio is evaluated for each active edge action and the loss is averaged over the active edges, so a micro-slot with more edges does not produce a larger loss. Because each policy changes the state distribution seen by the other, clipping and early stopping at a target Kullback--Leibler (KL) divergence limit the change per update.
\subsubsection{Additional Training Terms for the Manager}
All tasks admitted in the same macro-slot share the same advantage, although their splits may lead to different outcomes. The manager update therefore adds a credit term per task: once the outcomes are known, each task receives an adjustment relative to the mean outcome of its admission group, centered to zero mean so that the common reward is unchanged.

A second term, also used only in training, relies on the local feasibility feature $c_i$ of \eqref{eq:local_feasibility}. For tasks whose local execution is clearly infeasible or clearly feasible, the deterministic local share is pulled toward the corresponding
feasibility level:
\begin{equation}
\mathcal{L}_{\mathrm{cap},i}
=
\left(
\alpha_i^{\mathrm{det}}
-
\min(c_i,0.85)
\right)^2,
\label{eq:cap_anchor}
\end{equation}
which is applied only when
\begin{equation}
c_i\leq0.4
\qquad \text{or} \qquad
c_i\geq0.95.
\label{eq:cap_anchor_range}
\end{equation}
The intermediate range is left to RL. Both terms are used only during training.

Algorithm~\ref{alg:aeroglider} summarizes one training episode; the numerical hyperparameters are reported in Sec.~\ref{sec:setup}.

\begin{algorithm}[t]\small
\caption{AeroGLIDER Training Episode}
\label{alg:aeroglider}
\begin{algorithmic}[1]
\REQUIRE Manager parameters $\theta_{\mathrm{m}}$, worker parameters 
         $\theta_{\mathrm{w}}$, episode horizon $T$, energy budget $\overline{E}$,
         dual step $\eta_{\lambda}$, dual cap $\lambda_{\max}$, and current dual variable $\lambda$.
\STATE Reset the environment and rollout buffers.
\FOR{$t=1,\ldots,T$}
    \STATE Admit new tasks and construct $\mathcal{G}_{t}^{\mathrm{m}}$.
    \STATE Generate $(\alpha_i,\beta_i,\gamma_i)$ for each
            newly admitted task using the manager policy.
    \STATE Assign the resulting workloads to the local, UAV,
            and HAPS processing paths.
    \FOR{each $\tau\in\mathcal{K}_t$}
        \STATE Remove completed and deadline-expired tasks
             from the active sets.
        \STATE Construct worker graph $\mathcal{G}_{\tau}^{\mathrm{w}}$.
        \STATE Generate bandwidth, power, and CPU actions using the worker policy.
        \STATE Enforce the bandwidth and CPU constraints through the softmax of each resource, and the transmit power constraints through the projection.
        \STATE Apply the resource allocations and update
               workloads, queues, energy consumption, and task states.
        \STATE Record the completed work $\omega_{\tau}$ and worker transition.
    \ENDFOR
    \STATE Compute $r_t^{\mathrm{task}}$ and aggregate
        macro-slot energy $E_t$.
    \STATE Form the manager reward:
            $r_t^{\mathrm{m}}=r_t^{\mathrm{task}}-\lambda(E_t-\overline{E})$
    \STATE Assign $r_t^{\mathrm{m}}$ to the worker micro-slots
            using \eqref{eq:worker_reward}.
    \STATE Store the manager transition
    \STATE Update the dual variable: $\lambda\leftarrow\left[\lambda+\eta_{\lambda}
           (E_t-\overline{E})\right]_{0}^{\lambda_{\max}}$
\ENDFOR
\STATE Compute manager GAE over the $T$ macro-slot transitions.
\STATE Compute worker GAE over the $K_{\mathrm{slots}}T$ micro-slot transitions.
\STATE Associate each worker advantage with the active task-resource actions of its 
       micro-slot.
\STATE Update the manager actor and critic using clipped PPO, early stopping at the target KL, and the additional manager terms.
\STATE Update the worker actor and critic using clipped PPO and early stopping at the target KL.
\end{algorithmic}
\end{algorithm}

\subsubsection{Curriculum, Horizon Matching, and Selection}
\label{sec:curriculum}
\begin{itemize}
  \item \textbf{Curriculum over arrival rates.} Episodes are split
        $10\%/10\%/80\%$ into three phases that widen the sampled
        set of arrival rates $\mathbb{L}(e)$: Phase~1 trains on the lightest
        arrival rate only ($\mathbb{L}(e){=}\{240\}$), Phase~2 on the two
        lightest ($\{240,300\}$), and Phase~3 on the full set
        $\mathbb{L} = \{240, 300, \dots, 600\}$
        tasks/min (Sec.~\ref{sec:setup}); each episode draws $\Lambda$ uniformly
        from $\mathbb{L}(e)$. The easy phases are kept short so
        that $80\%$ of the training budget is spent on the full
        mix, which gives the most exposure to the highest arrival rates
        ($540$--$600$ tasks/min).
  \item \textbf{Horizon matching.} Training episodes span the
        same horizon $T$ as the evaluation simulator.
  \item \textbf{Checkpoint tracking and selection.} During training, the success rate averaged over the last $25$ episodes is tracked, and a checkpoint is saved whenever it reaches a new best (its history is reset when Phase~3 starts, so a policy specialized to light traffic is never locked in). Because this rolling success rate is measured with sampled actions and can disagree with deterministic deployment, the deployed parameters $\theta^\star$ are then chosen among the rolling best, snapshots at fixed episodes, and the final checkpoint by deterministic evaluation on the validation seeds (Sec.~\ref{sec:setup}).
\end{itemize}

\subsection{Computational Complexity}
\label{sec:complexity}
The computational cost of AeroGLIDER is mainly determined by the heterogeneous message-passing layers and the actor heads. Let $d_h$ denote the hidden embedding dimension, $L$ the number of message-passing layers, and $H$ the number of attention heads. For a graph $\mathcal{G}=(\mathcal{V},\mathcal{E})$, one forward pass through the GATv2 encoder has complexity
\begin{equation}
\mathcal{O}
\left(
L
\left(
|\mathcal{V}|d_h^2
+
H|\mathcal{E}|d_h
\right)
\right),
\label{eq:gnn_complexity}
\end{equation}
where the first term is the transformation of the node features and the second the attention over the edges. Since $L$, $H$, and $d_h$ are fixed, the inference cost grows linearly with the number of nodes and edges.

For the manager graph, let $n_t^{\mathrm{m}}=|\mathcal{B}(t)|$ denote the number of newly admitted tasks. Its node count is
\begin{equation}
|\mathcal{V}_t^{\mathrm{m}}|
=
n_t^{\mathrm{m}}+M+1.
\end{equation}
so the manager cost grows linearly with the number of admitted tasks.

For the worker graph, let $n_{\tau}^{\mathrm{w}}=
|\mathcal{V}_{\tau}^{\mathrm{task}}|$ denote the number of pending tasks. Its node count is \begin{equation}
|\mathcal{V}_{\tau}^{\mathrm{w}}|
=
n_{\tau}^{\mathrm{w}}
+
(M+1)
+
(N+M+1).
\end{equation}
and $|\mathcal{E}_{\tau}^{\mathrm{w}}|$ grows linearly with $n_{\tau}^{\mathrm{w}}$ (the reverse relation only changes the constant). The actor heads are evaluated once per admitted task or active edge and the critics pool node embeddings, so with fixed network dimensions the overall inference complexity is
\begin{equation}
\mathcal{O}
\left(
|\mathcal{V}|+|\mathcal{E}|
\right)
\end{equation}
with respect to the instantaneous graph size.
Since the parameters are shared across tasks and edges, a heavier workload increases the number of actor evaluations but not the number of learned parameters. Graph construction and the power projection add an overhead per step that also grows with the active graph; the measurements of Sec.~\ref{sec:runtime} include them.

\section{Simulation Results and Performance Evaluation}
\label{sec:results}
The experiments examine the success rate under increasing task arrival rate, the contribution of the main components, zero-shot generalization, and the inference cost at deployment against DRL baselines and a receding horizon linear programming (RH-LP) reference.

\subsection{Simulation and Training Setup}
\label{sec:setup}
\subsubsection{Network and Task Configuration}
We evaluate AeroGLIDER in the HAPS-UAV-IoT system of Sec.~\ref{sec:system_model}, with $N=100$ IoT devices, $M=4$ UAVs, and one HAPS, i.e., one cluster of four UAV cells served by one HAPS, the scale of the HALO evaluation~\cite{cai2026prior}. Five placement configurations are used. In each, the devices follow four Gaussian traffic hotspots over an area of about $0.7\times0.7$~km, the truncated Gaussian hotspot model commonly used for clustered ground users~\cite{mozaffari2017hovertime}, and each UAV is placed above a $k$-means centroid of the device locations, a standard placement rule driven by demand~\cite{galkin2016uavdeploy}. Across the five configurations, $14$--$54\,\%$ of the devices lie in the overlap of two coverage disks, so the association auction has a real choice. The UAV altitude of $100$~m stays below the $120$~m ceiling for small unmanned aircraft~\cite{mozaffari2019uavtutorial}, the coverage radius of $220$~m keeps the elevation angle at the cell edge at $24^{\circ}$, where the urban LoS probability of~\eqref{eq:plos} is still above one half~\cite{alhourani2014los}, and the HAPS is $20$~km above the center of the area, the nominal stratospheric operating altitude~\cite{kurt2021hapsmag,3gpp38811}. The serving UAV of each task is chosen by the auction of Sec.~\ref{sec:association}.

Tasks arrive according to a Poisson process with aggregate arrival rate $\Lambda$ (Sec.~\ref{sec:traffic}), which we express in tasks per minute. The training range is
\begin{equation}
\Lambda\in\{240,300,360,420,480,540,600\}
\quad\text{tasks/min},
\label{eq:training_loads}
\end{equation}
which corresponds to $4$--$10$ tasks/s on average. Each task belongs to a small, medium, or large class with probabilities $0.6$, $0.3$, and $0.1$; the size and deadline ranges of each class are given in Table~\ref{tab:sim_params}. The micro-slot duration is $\Delta t=50$~ms, with $K_{\mathrm{slots}}=20$ micro-slots per $1$-s macro-slot; all remaining system and task parameters are listed in Table~\ref{tab:sim_params}.

\begin{table}
\caption{Simulation parameters and the basis of each value: a supporting reference, $\dagger$ a design choice motivated in Sec.~\ref{sec:setup_basis}, or $\ddagger$ a value derived from other entries.}
\label{tab:sim_params}
\centering
\footnotesize
\adjustbox{max width=\columnwidth}{%
\begin{tabular}{@{}lll@{\hskip 5pt}l@{}}
\toprule
\textbf{Parameter} & \textbf{Symbol} & \textbf{Value} & \textbf{Basis} \\ \midrule
Micro-slot duration         & $\Delta t$            & $50$~ms & $\dagger$ \\
Micro-slots per macro-slot  & $K_{\mathrm{slots}}$  & $20$ & $\ddagger$ \\
Macro-slot duration         & $\Delta T$            & $1$~s & $\dagger$ \\
Episode length              & $T$                   & $600$~s & $\dagger$ \\
IoT devices                 & $N$                   & $100$ & \cite{cai2026prior} \\
UAVs                        & $M$                   & $4$ & \cite{cai2026prior} \\
UAV altitude                & $H_m$                 & $100$~m & \cite{mozaffari2019uavtutorial} \\
UAV coverage radius         & $R_{\mathrm{cov}}$    & $220$~m & $\ddagger$, \cite{alhourani2014los} \\
HAPS altitude               & $H_{\mathrm{H}}$      & $20$~km & \cite{kurt2021hapsmag,3gpp38811} \\
IoT transmit power          & $P_{\max}^{\mathrm{IoT}}$      & $0.2$~W ($23$~dBm) & \cite{3gpp36101,3gpp38821} \\
UAV transmit power          & $P_{\max}^{\mathrm{UAV}}$      & $2.0$~W ($33$~dBm) & \cite{3gpp38821,3gpp36104} \\
IoT compute rate            & $F_{\max}^{\mathrm{IoT}}$      & $1.0$~Gcyc/s & \cite{chen2016multiuser,mao2016dynamic} \\
UAV compute rate            & $F_{\max}^{\mathrm{UAV}}$      & $4.0$~Gcyc/s & $\dagger$ \\
HAPS compute rate           & $F_{\max}^{\mathrm{HAPS}}$     & $10.0$~Gcyc/s & $\dagger$ \\
Compute intensity           & $C$                   & $800$~cycles/bit & $\dagger$, \cite{mao2016dynamic} \\
Task classes (S/M/L)        & --                    & $60/30/10\,\%$ & $\dagger$ \\
Task size per class         & $D_i$                 & $\mathcal{U}(0.5,4)/(4,20)$ & $\dagger$, \cite{chen2016multiuser,jiaHierarchicalAerialComputing2023} \\
  & & $/(20,80)$~Mbit & \\
Deadline per class          & $T_i^{\max}$          & $\mathcal{U}(2,10)/(2.5,15)$ & $\dagger$ \\
   & & $/(8,20)$~s & \\
IoT$\to$UAV bandwidth       & $W_{\mathrm{IU}}$     & $3$~MHz & \cite{3gpp36101} \\
UAV$\to$HAPS bandwidth      & $W_{\mathrm{UH}}$     & $5$~MHz & \cite{3gpp38811} \\
Carrier frequency           & $f_c$                 & $2.0$~GHz & \cite{3gpp38811,alhourani2014los} \\
Path-loss exp.\ (A2A)       & $\alpha_{AA}$         & $2.0$ & \cite{goldsmith2005wireless} \\
A2A antenna gain            & $G_{\mathrm{A2A}}$    & $40$~dB & \cite{3gpp38821} \\
A2G LoS params (urban)      & $a,b$                 & $9.61,\,0.16$ & \cite{alhourani2014los} \\
LoS/NLoS excess loss        & $\eta_{\mathrm{LoS}},\eta_{\mathrm{NLoS}}$ & $1,\,20$~dB & \cite{alhourani2014pathloss} \\
Noise PSD ($kT$ $+$ $7$~dB NF) & $N_0$              & $-167$~dBm/Hz & \cite{goldsmith2005wireless,3gpp38821} \\
CMOS coefficient            & $\kappa$              & $10^{-28}$ & \cite{burd1996dvfs,mao2016dynamic} \\
Deployment configurations   & --                    & $5$ & $\dagger$ \\
Per-step energy budget      & $\overline{E}$        & $15$~J/step ($=15$~W) & $\ddagger$ \\
\bottomrule
\end{tabular}}
\end{table}

\subsubsection{Basis of the Simulation Settings}
\label{sec:setup_basis}
The physical-layer parameters in Table~\ref{tab:sim_params}
follow commonly used 3GPP NTN and aerial-link settings, including
the $2$~GHz carrier, $23$~dBm IoT transmit power, A2G propagation
parameters, and UAV-HAPS link configuration. The computation
capacities are selected to represent progressively stronger local,
UAV-edge, and HAPS tiers, while the task-size and deadline ranges
span workloads for which different execution tiers are beneficial.
The $50$-ms micro-slot accommodates the measured worker inference
latency in Sec.~\ref{sec:runtime}, and the $1$-s macro-slot matches
the task-arrival timescale. The training range
$\Lambda=240$-$600$ tasks/min spans moderate to strongly
oversubscribed operating conditions; higher arrival rates are
reserved for the zero-shot tests of Sec.~\ref{sec:generalization}.

\subsubsection{Training Configuration}
AeroGLIDER is implemented in PyTorch~\cite{paszke2019pytorch} using PyTorch Geometric~\cite{fey2019pyg}, and consists of a hidden dimension of $128$, two heterogeneous GATv2 message-passing layers, and four attention heads. The manager
and worker are trained with PPO using the hyperparameters in Table~\ref{tab:hype}. Standard PPO parameters follow established settings~\cite{schulman2017ppo,engstrom2020implementation}, while AeroGLIDER-specific coefficients were selected using validation traffic on a deployment excluded from the evaluation. Test seeds were not used for hyperparameter selection.  Each training run contains $300$ episodes of the $600$-s evaluation horizon and follows the curriculum and the checkpoint selection of Sec.~\ref{sec:curriculum}; the test traffic is never used for checkpoint selection. All training runs use a single NVIDIA A100 GPU with $40$~GB of memory.

\begin{table}
\caption{AeroGLIDER hyperparameters}
\label{tab:hype}
\begin{tabular}{ll}
\toprule
\textbf{Component} & \textbf{Value} \\ \midrule
Hidden dim ($d_h$)              & $128$ \\
GATv2 message-passing layers    & $2$ \\
GATv2 attention heads           & $4$ \\
\multicolumn{2}{l}{Feature scales $\rho_z$ of~\eqref{eq:feature_scaling}:} \\
\quad task size / distance      & $10$~Mbit / $100$~m \\
\quad urgency                   & $2$~Mbit/s \\
\quad UAV backlog (tx / comp)   & $50$~Mbit / $10$~Gcycle \\
\quad HAPS backlog (tx / comp)  & $50$~Mbit / $50$~Gcycle \\
\quad workload (tx / comp)      & $1$~Mbit / $1$~Gcycle \\
\quad queue depth               & $10$ \\
Worker learning rate            & $3 \times 10^{-4}$ \\
Manager learning rate           & $2 \times 10^{-4}$ \\
Discount $\gamma$ (worker/mgr)  & $0.95$ / $0.99$ \\
GAE parameter $\lambda_{\mathrm{GAE}}$ & $0.95$ \\
PPO clip $\epsilon$             & $0.2$ \\
PPO epochs                      & $10$ \\
Target KL                       & $0.02$ \\
Entropy coefficient (worker / manager) & $0.03$ / $0.003$ \\
Manager log-std init            & $-1.0$ \\
Dual variable ($\lambda_0$ / $\eta_\lambda$ / $\lambda_{\max}$) & $0.1$ / $10^{-4}$ / $10$ \\
Task credit weight (per task)   & $1.0$ \\
Feasibility anchor weight / band & $4.0$ / $[0.4, 0.95]$ \\
Minibatch size                  & $512$ \\
Worker update interval          & $2400$ steps \\
Manager update interval         & $240$ steps \\
Curriculum (Phase 1/2/3)        & $10\%$ / $10\%$ / $80\%$ \\
Total episodes (all methods)    & $300^{\dagger}$ \\
Optimizer                       & Adam~\cite{kingma2015adam} \\
Learning rate schedule          & Cosine, $5\%$ warmup,\\
 &floor $0.2{\times}$ \\
\bottomrule
\end{tabular}
\vspace{1mm}
\parbox{\linewidth}{%
\scriptsize
$^{\dagger}$ One training run per method and deployment configuration. Checkpoints are selected using validation seeds only.
}
\end{table}

\subsubsection{Baselines and RH-LP Reference}
\label{sec:baselines}
We compare AeroGLIDER with three standard DRL baselines: PPO~\cite{schulman2017ppo},
DDPG~\cite{lillicrap2016ddpg}, and SAC~\cite{haarnoja2018sac}. All three use fully connected networks with two hidden layers of $256$ neurons and the same flat state and action interface. Their hyperparameters and training configurations are summarized in Table~\ref{tab:baseline_hyp}.

The flat interface holds at most $n_{\max}=10$ new tasks per macro-slot, the mean arrival count at the highest training rate. Further arrivals cannot be represented and are dropped at admission, and these drops count as failed tasks. Pending tasks are ordered by remaining slack, the policy controls the first $n_{\max}$ of each queue, and the rest wait in the simulator until they enter the controlled window.

Each flat baseline outputs an $n_{\max}\times8$ action matrix per macro-slot ($80$ outputs): for each task slot, the UAV and HAPS fractions $(\beta_i,\gamma_i)$ with $\alpha_i=1-\beta_i-\gamma_i$, and the transmit power, bandwidth, and CPU controls of the access and backhaul links. The flat policies act once per $1$-s macro-slot and their allocations are reused over the $K_{\mathrm{slots}}=20$ micro-slots. For fairness, PPO, DDPG, and SAC use the same arrival rates, episode horizon, training budget, curriculum, deployment configurations, and checkpoint selection on validation seeds as AeroGLIDER.

We also include a \emph{receding horizon linear programming (RH-LP) reference}. At each macro-slot it observes the full system state (backlogs, residual workloads, deadlines), solves a linear program over the completion fractions and tier assignments of the tasks under the comm., computation, and energy limits, executes only the decisions of the current macro-slot, and solves again at the next one; it does not know the future arrivals. Its decisions run in the same simulator as AeroGLIDER. Since RH-LP needs an explicit system model and repeated online optimization, it serves as an achievable model-based reference, not as an optimum.

\begin{table}[!tp]
\centering
\caption{Training configurations of the DRL baselines.}
\label{tab:baseline_hyp}
\renewcommand{\arraystretch}{1.02}
\setlength{\tabcolsep}{4pt}
\footnotesize
\adjustbox{max width=\columnwidth}{%
\begin{tabular}{lccc}
\toprule
 & \textbf{PPO} & \textbf{DDPG} & \textbf{SAC} \\
\midrule
Learning rate(s) $\times 10^{-4}$ & $3$ & $0.5$ / $1$ & $1$ / $1$ \\
Discount $\gamma$    & $0.99$ & $0.99$ & $0.99$ \\
Advantage / target   & MC return & $\tau{=}0.002$ & $\tau{=}0.002$ \\
PPO clip $\epsilon$ / epochs & $0.2$ / $20$ & -- & -- \\
Update cadence (steps) & $2000$ & every $2$ & every $2$ \\
Replay size / batch  & -- & $10^{5}$ / $256$ & $10^{5}$ / $256$ \\
Exploration          & ent.\ $0.02$ & noise $0.08{\to}0.005$ & $\alpha{=}0.03$ \\
Action filter (train $=$ deploy) & -- & $0.90$ & $0.90$ \\
\bottomrule
\end{tabular}}
\end{table}

\subsubsection{Evaluation Protocol and Metrics}
\label{sec:evaluation_protocol}
Each method is trained once per deployment configuration (five runs per method, fixed training seed). The selected checkpoint of each run is evaluated deterministically on ten test traffic seeds per arrival rate, disjoint from the training and validation traffic, so each reported point averages $50$ episodes unless stated otherwise; we report means with $95\%$ confidence intervals.

The primary metric is success rate,
$\mathrm{SR}=N_{\mathrm{succ}}/N_{\mathrm{arr}}\times100\%$.
For the flat baselines we additionally report the interface-induced admission-drop rate
$R_{\mathrm{drop}}=N_{\mathrm{drop}}^{\mathrm{int}}/
N_{\mathrm{arr}}\times100\%$
and success rate over admitted tasks
$\mathrm{SR}_{\mathrm{adm}}=
N_{\mathrm{succ}}/N_{\mathrm{adm}}\times100\%$.
We further report the IoT and UAV energy per macro-slot and per delivered megabit. Since only AeroGLIDER is trained under the energy constraint $\overline{E}$, these values are the operating points reached by each method rather than a constrained comparison (Sec.~\ref{sec:discussion}).

\subsection{Main Performance Comparison}
\label{sec:main_perf}
We first evaluate AeroGLIDER at the seven training arrival rates of~\eqref{eq:training_loads}, whose spacing and end points are motivated in Sec.~\ref{sec:setup_basis}. 
\subsubsection{Success Rate Under Increasing Task Arrival Rate}
\begin{figure*}[!t]
\centering
\includegraphics[width=0.98\textwidth]{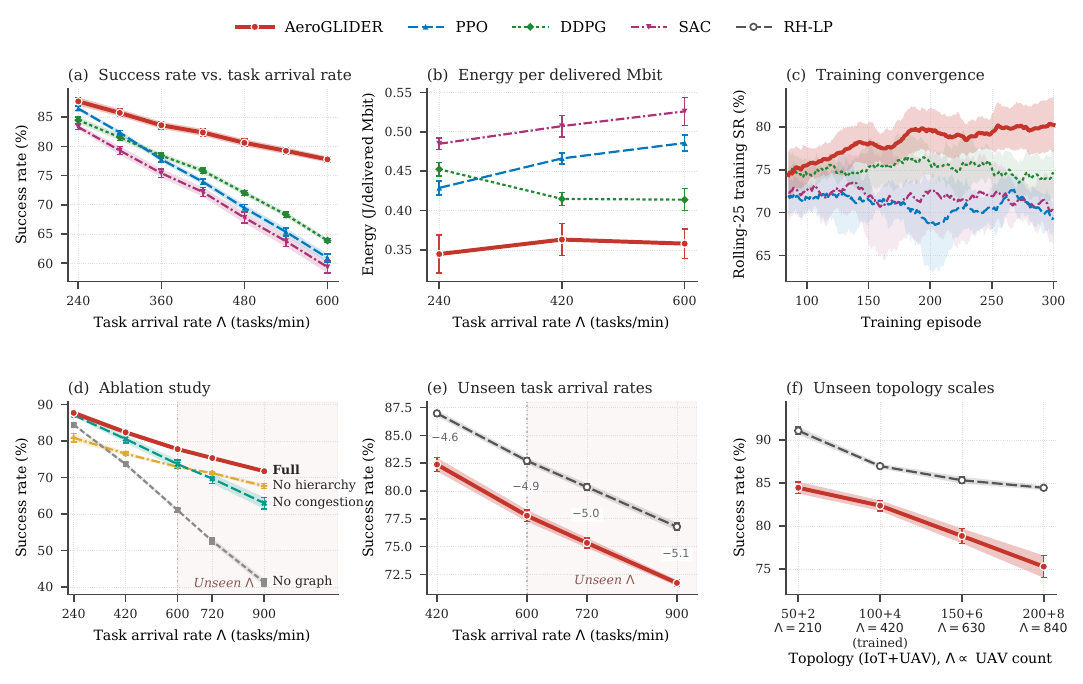}
\caption{Performance evaluation of AeroGLIDER.
(a) Success rate vs.\ task arrival rate over the training range.
(b) Energy per delivered Mbit.
(c) Training convergence during the final curriculum phase.
(d) Ablation study.
(e) Unseen task arrival rates, compared with the RH-LP reference.
(f) Unseen topology scales, compared with the RH-LP reference.
Shaded regions in (d) and (e) mark the unseen arrival rates beyond the training maximum $\Lambda{=}600$. Results are averaged over five deployment configurations and ten test seeds unless otherwise stated. Error bars and shaded
bands denote $95\%$ confidence intervals.}
\label{fig:panorama}
\end{figure*}

Fig.~\ref{fig:panorama}(a) and Table~\ref{tab:main_results} compare the task success rate (SR) as the arrival rate increases. All methods lose SR under heavier traffic, but AeroGLIDER degrades far more slowly: its SR falls from $87.7\%$ at $\Lambda=240$ to $77.8\%$ at $\Lambda=600$ ($9.9$ percentage points, pp), whereas PPO, DDPG, and SAC lose $20.6$-$25.6$ pp, so the margin over the strongest baseline widens from $1.2$ pp to $13.9$ pp. The margin is smallest when resources are lightly contested and largest when many tasks are active at the same time, which is consistent with a graph whose size follows the workload and with the reallocation of resources at every micro-slot; the ablation study in Sec.~\ref{sec:ablation} isolates these contributions.

\subsubsection{Interface Scalability and Decision Quality}
Part of the overall gap is structural: as the arrival rate increases, more macro-slots contain more than $n_{\max}=10$ new tasks, and the admission drop rate of the flat baselines rises from $0.1\%$ at $\Lambda=240$ to $12.5\%$ at $\Lambda=600$ (Table~\ref{tab:main_results}), whereas AeroGLIDER admits every arrival. Table~\ref{tab:main_results} therefore also reports the baseline SR over admitted tasks only. Removing the admission drops narrows the gap but does not close it: at $\Lambda=600$ the margin over DDPG falls from $13.9$ pp overall to $4.7$ pp over admitted tasks, so the split and allocation decisions of AeroGLIDER contribute beyond the interface effect.

\begin{table*}[!tp]
\centering
\caption{Success rate, interface-induced admission drops, and
admitted-task success rate across task arrival rates
(mean $\pm95\%$ CI over $50$ test episodes).}
\label{tab:main_results}
\renewcommand{\arraystretch}{0.95}
\setlength{\tabcolsep}{7pt}
\footnotesize
\begin{tabular}{l *{7}{c}}
\toprule
& \multicolumn{7}{c}{Task arrival rate $\Lambda$ (tasks/min)} \\
\cmidrule(lr){2-8}
\textbf{Method} & 240 & 300 & 360 & 420 & 480 & 540 & 600 \\
\midrule
\multicolumn{8}{l}{\emph{Success rate over all arrivals (\%)}} \\
PPO      & 86.5\,{\scriptsize$\pm$0.3} & 82.4\,{\scriptsize$\pm$0.4} & 77.8\,{\scriptsize$\pm$0.4} & 73.9\,{\scriptsize$\pm$0.5} & 69.5\,{\scriptsize$\pm$0.6} & 65.4\,{\scriptsize$\pm$0.7} & 60.9\,{\scriptsize$\pm$0.7} \\
DDPG     & 84.5\,{\scriptsize$\pm$0.6} & 81.6\,{\scriptsize$\pm$0.5} & 78.4\,{\scriptsize$\pm$0.4} & 75.8\,{\scriptsize$\pm$0.5} & 72.1\,{\scriptsize$\pm$0.4} & 68.3\,{\scriptsize$\pm$0.4} & 63.9\,{\scriptsize$\pm$0.4} \\
SAC      & 83.3\,{\scriptsize$\pm$0.4} & 79.3\,{\scriptsize$\pm$0.7} & 75.4\,{\scriptsize$\pm$0.7} & 72.2\,{\scriptsize$\pm$0.8} & 67.8\,{\scriptsize$\pm$0.9} & 63.8\,{\scriptsize$\pm$0.9} & 59.4\,{\scriptsize$\pm$1.0} \\
\textbf{AeroGLIDER} & \textbf{87.7}\,{\scriptsize$\pm$0.7} & \textbf{85.7}\,{\scriptsize$\pm$0.6} & \textbf{83.6}\,{\scriptsize$\pm$0.6} & \textbf{82.4}\,{\scriptsize$\pm$0.6} & \textbf{80.6}\,{\scriptsize$\pm$0.6} & \textbf{79.2}\,{\scriptsize$\pm$0.6} & \textbf{77.8}\,{\scriptsize$\pm$0.5} \\
\midrule
Admission drop rate of the flat interface (\%) & 0.1\,{\scriptsize$\pm$0.0} & 0.4\,{\scriptsize$\pm$0.0} & 1.4\,{\scriptsize$\pm$0.1} & 2.8\,{\scriptsize$\pm$0.1} & 5.3\,{\scriptsize$\pm$0.1} & 8.7\,{\scriptsize$\pm$0.2} & 12.5\,{\scriptsize$\pm$0.2} \\
\midrule
\multicolumn{8}{l}{\emph{Success rate over admitted tasks only (\%; flat baselines)}} \\
PPO      & 86.6\,{\scriptsize$\pm$0.3} & 82.7\,{\scriptsize$\pm$0.4} & 78.9\,{\scriptsize$\pm$0.4} & 76.1\,{\scriptsize$\pm$0.5} & 73.4\,{\scriptsize$\pm$0.6} & 71.6\,{\scriptsize$\pm$0.7} & 69.6\,{\scriptsize$\pm$0.8} \\
DDPG     & 84.6\,{\scriptsize$\pm$0.6} & 81.9\,{\scriptsize$\pm$0.5} & 79.5\,{\scriptsize$\pm$0.4} & 78.0\,{\scriptsize$\pm$0.5} & 76.1\,{\scriptsize$\pm$0.4} & 74.8\,{\scriptsize$\pm$0.4} & 73.0\,{\scriptsize$\pm$0.4} \\
SAC      & 83.4\,{\scriptsize$\pm$0.4} & 79.6\,{\scriptsize$\pm$0.7} & 76.5\,{\scriptsize$\pm$0.7} & 74.3\,{\scriptsize$\pm$0.8} & 71.6\,{\scriptsize$\pm$1.0} & 69.8\,{\scriptsize$\pm$1.0} & 67.8\,{\scriptsize$\pm$1.1} \\
\bottomrule
\end{tabular}
\end{table*}

\subsubsection{Energy Consumption}
AeroGLIDER consumes more energy per macro-slot than the flat baselines because it completes more of the offered workload: $9.1$-$12.9$~J/step from $\Lambda=240$ to $600$, against $8.8$-$9.1$ and $10.7$-$11.6$~J/step for the baselines. Fig.~\ref{fig:panorama}(b) therefore normalizes energy by the delivered workload. AeroGLIDER uses the least energy per delivered Mbit at all three evaluated arrival rates, about $0.34$-$0.36$~J/Mbit against $0.41$-$0.53$~J/Mbit for the baselines. We do not claim an absolute energy advantage, since only AeroGLIDER is trained under $\overline{E}$; the results show that it stays within its energy budget on average while serving substantially more of the offered workload.

\subsubsection{Work Distribution Across Computing Tiers}
Table~\ref{tab:shares} reports the workload fractions the manager routes to local, UAV, and HAPS execution. All three tiers remain in use at every arrival rate, so the manager does not collapse to a single tier, and the local share rises modestly ($45.0\%$ to $48.9\%$) as contention on the shared aerial resources grows. How the split depends on the task at unseen arrival rates is examined in Sec.~\ref{sec:generalization}.

\begin{table}[!tp]
\centering
\caption{AeroGLIDER manager workload distribution across computing tiers under increasing task arrival rate.}
\label{tab:shares}
\renewcommand{\arraystretch}{0.95}
\setlength{\tabcolsep}{4.5pt}
\footnotesize
\begin{tabular}{l *{7}{c}}
\toprule
 & \multicolumn{7}{c}{Task arrival rate $\Lambda$ (tasks/min)} \\
\cmidrule(lr){2-8}
\textbf{Tier} & 240 & 300 & 360 & 420 & 480 & 540 & 600 \\
\midrule
Local & 45.0 & 45.4 & 45.7 & 46.3 & 47.2 & 47.9 & 48.9 \\
UAV   & 20.7 & 20.7 & 20.6 & 20.3 & 19.8 & 19.4 & 19.0 \\
HAPS  & 34.3 & 33.9 & 33.8 & 33.4 & 33.0 & 32.7 & 32.1 \\
\bottomrule
\end{tabular}
\end{table}

\subsection{Training Convergence}
\label{sec:convergence}
Fig.~\ref{fig:panorama}(c) reports the success rate averaged over the last $25$ episodes during Phase~3 of the curriculum (Sec.~\ref{sec:curriculum}), the only phase in which the set of arrival rates is fixed; in Phases~1--2 the success rate depends mainly on the sampled arrival rate rather than on policy improvement.

The flat baselines plateau early in Phase~3, at about $71\%$ (PPO), $72\%$ (SAC), and $75\%$ (DDPG), whereas AeroGLIDER keeps improving from about $74\%$ to about $80\%$, overtaking the best baseline around episode~90. The flat policies thus gain little once the full range of arrival rates is introduced, while AeroGLIDER keeps refining its split and allocation decisions.

The curves are measured with sampled actions (PPO, AeroGLIDER) or filtered exploration actions (DDPG, SAC); they indicate learning progress and do not reproduce the deterministic test results.

\subsection{Ablation Studies}
\label{sec:ablation}
We evaluate three ablation variants to examine the contribution of the main design components of AeroGLIDER: the graph representation, the hierarchical control structure, and the congestion information in the state. Each variant removes one component from the complete framework and is retrained with the training and checkpoint selection procedure of Sec.~\ref{sec:setup}.
\paragraph{No Graph Representation}
The \emph{No Graph} variant replaces the manager and worker graphs with the flat interface of the baselines, including $n_{\max}=10$ admission slots and queue windows of $10$ tasks, while keeping the two-timescale control, reward, energy constraint, and training procedure of AeroGLIDER. Since this slot interface is the one used by HALO~\cite{cai2026prior}, the variant also serves as a flat hierarchical reference in the style of HALO, trained under the same conditions as AeroGLIDER.

\paragraph{No Hierarchy}
The \emph{No Hierarchy} variant keeps the graph representations and network capacity but removes the split into manager and worker. The two decision levels are instead trained as a single PPO process with one reward, one discount factor, and one optimization stream.

\paragraph{No Congestion Awareness}
The \emph{No Congestion} variant keeps both the graph structure and the hierarchy but removes the congestion inputs: the resource backlogs and queue depths of the worker and the backlog features of the manager. The manager guidance that depends on congestion is also disabled.

Fig.~\ref{fig:panorama}(d) compares the complete AeroGLIDER
framework with the three ablation variants. The complete model achieves success rates of $87.7\%$, $82.4\%$, and $77.8\%$ at $\Lambda=240$, $420$, and $600$ tasks/min, respectively. All three ablations reduce performance, but they behave differently as the arrival rate increases. Removing the graph interface causes the largest loss under heavy traffic: relative to the complete model, the success rate gap grows from $3.2$ pp at $\Lambda=240$ to $16.7$ pp at $\Lambda=600$. This widening gap is consistent with the flat interface becoming more restrictive as the number of active tasks grows.

Removing the hierarchy causes a large loss over the whole range of arrival rates: the success rate gap decreases from $6.8$ pp at $\Lambda=240$ to $4.9$ pp at $\Lambda=600$. The two-timescale decomposition thus improves task performance.
The differences of the No Graph and No Congestion variants grow outside the training range. At $\Lambda=900$ tasks/min, the complete model keeps a success rate of $71.7\%$, while the No Hierarchy variant falls to $67.6\%$, a gap of $4.1$ pp. The No Congestion variant reaches $63.0\%$, a gap of $8.7$ pp, and the No Graph variant falls to $41.3\%$.
Overall, the ablations support three observations. The graph interface matters most when many tasks are active, the hierarchical decomposition improves performance, and explicit congestion information becomes more valuable as resource contention grows.

\subsection{Generalization}
\label{sec:generalization}
All experiments in this subsection reuse the selected checkpoints without retraining and consider higher task arrival rates and different network scales.

\subsubsection{Generalization to Unseen Task Arrival Rates}
AeroGLIDER is trained with arrival rates up to $\Lambda=600$ tasks/min. We evaluate the same deployed policies at $\Lambda=720$ and $\Lambda=900$ tasks/min, i.e., $20\%$ and $50\%$ above the maximum training arrival rate.
Fig.~\ref{fig:panorama}(e) compares AeroGLIDER with the RH-LP reference at both the trained and the unseen arrival rates. The mean success rate of AeroGLIDER decreases from $77.8\%$ at the highest training arrival rate to $75.3\%$ at $\Lambda=720$ and $71.7\%$ at $\Lambda=900$, a further loss of only $6.0$ pp for $50\%$ more traffic than seen in training.

The gap to the RH-LP reference stays at $4.6$-$4.9$ pp over the training range and $5.0$-$5.1$ pp at the two unseen rates, so the loss follows that of the model-based reference rather than diverging from it. The manager also keeps adapting the split to each task at $\Lambda=900$: the correlation between task size and local fraction is $\rho=-0.86$, i.e., larger tasks are shifted toward the shared UAV and HAPS resources instead of collapsing to a fixed split.
\subsubsection{Generalization to Unseen Network Scales}
The same checkpoints, trained on $100$ devices and $4$ UAVs, are evaluated on three unseen topologies, $(50,2)$, $(150,6)$, and $(200,8)$ devices and UAVs, i.e., $0.5\times$ to $2\times$ the training scale at the same $25$ devices per UAV.
To keep the traffic per UAV about the same, the arrival rate is scaled with the number of UAVs as $\Lambda=105\times M$ tasks/min. This gives $\Lambda=210$, $630$, and $840$ tasks/min for the three unseen topologies, while the original topology with $100$ devices and $4$ UAVs is evaluated at the matching $\Lambda=420$. Device locations are regenerated and the association auction is repeated for each topology.
Fig.~\ref{fig:panorama}(f) shows that the unchanged weights keep working at all scales: the mean success rate is $84.5\%$ with $50$ devices and $2$ UAVs, $82.4\%$ on the training topology at the matching rate, and $75.3\%$ with $200$ devices and $8$ UAVs. The transfer is possible because a different network size changes only the graph, not the learned parameters; the flat baselines, whose layers are tied to the training topology, cannot be applied without retraining.

\begin{table}[!tp]
\centering
\caption{Model size and inference latency of the AeroGLIDER manager and worker at deployment.}
\label{tab:deploy}
\renewcommand{\arraystretch}{1.04}
\setlength{\tabcolsep}{5pt}
\footnotesize
\begin{tabular}{lcc}
\toprule
 & \textbf{Manager} & \textbf{Worker} \\
\midrule
Decision rate               & $1$~Hz & $20$~Hz \\
Latency budget per decision & $1$~s  & $50$~ms \\
Parameters                  & $318.7$K & $327.5$K \\
Checkpoint size on disk     & \multicolumn{2}{c}{$2.6$~MB total} \\
Input graph nodes           & arrivals ${+}\,5$ & pending tasks ${+}\,110$ \\
Graph construction (CPU)    & --- & $0.3$--$0.4$~ms \\
Forward, $8$-thread Xeon CPU & $3.9$--$4.1$~ms & $4.5$--$5.1$~ms \\
Forward, NVIDIA A100        & $5.5$~ms & $6.0$--$6.4$~ms \\
Flat baseline forward (CPU, ref.) & \multicolumn{2}{c}{${\sim}0.1$~ms} \\
\bottomrule
\end{tabular}
\end{table}
\subsection{Scalability and Inference Cost}
\label{sec:runtime}
Table~\ref{tab:deploy} summarizes the deployment profile of the manager and worker: decision rates, latency budgets, parameter counts ($646$K in total, $2.6$~MB), graph sizes, and measured CPU/GPU inference latencies.
The near-equal parameter counts reflect two design assumptions. Both policies run on a single controller that observes the queue states of all three tiers, and they share the same encoder configuration ($d_h=128$, two GATv2 layers, four attention heads), so their parameter counts are set by the hidden dimension and the number of node and relation types, not by the decision rate. What differs by a factor of $20$ is the number of decisions per second and hence the latency budget per decision. The cost of one decision is set by the input graph: the manager graph holds only the tasks admitted in the current macro-slot plus $M+1$ resource nodes, whereas the worker graph holds all pending tasks plus $N+2M+2$ resource nodes. The $50$-ms budget of the worker is therefore the binding one. It is met with a margin of about nine, so the $20$ worker decisions of a macro-slot use about $11\,\%$ of the evaluated CPU, while the manager uses less than $0.5\,\%$ of its $1$-s budget.
On an $8$-thread Xeon CPU, manager inference takes $3.9$-$4.1$~ms against its $1$-s budget, and the complete worker decision (graph construction plus forward pass) averages $4.8$-$5.5$~ms against the $50$-ms micro-slot. CPU inference is slightly faster than the A100 GPU at these graph sizes, so GPU acceleration is not required.

\subsection{Scope and Limitations}
\label{sec:discussion}
Three points define the scope of the evaluation. First, AeroGLIDER is trained under the average energy constraint $\overline{E}$ whereas the DRL baselines are not, so the reported energy values are the operating points reached by each method rather than a constrained comparison, and the single Lagrange multiplier regulates energy over the training distribution rather than for each arrival rate. Second, the device-UAV association is fixed by the auction of Sec.~\ref{sec:association} and is not learned. Third, the experiments use static deployments, synthetic Poisson traffic, and the statistical channel model of Sec.~\ref{sec:system_model}; mobility and measured channels are outside the present scope, although the zero-shot tests of Sec.~\ref{sec:generalization} cover unseen arrival rates and network scales.

\section{Conclusion}
\label{sec:conclusion}
This paper proposed \emph{AeroGLIDER}, a hierarchical DRL framework on heterogeneous graphs for joint task offloading and resource allocation in HAPS-assisted HAPS-UAV-IoT networks. A manager splits each task across local, UAV, and HAPS execution once per macro-slot, and a worker reallocates communication and computation resources at every micro-slot. A graph whose size follows the workload, GATv2 message passing, critics on the graph, and hierarchical PPO let the control policy adapt to changing task populations and resource congestion without a predefined maximum task count.

Across different task arrival rates and deployment configurations, AeroGLIDER achieves higher task success rates than PPO, DDPG, and SAC, with the margin over the strongest baseline increasing from $1.2$ to $13.9$ percentage points as the arrival rate rises from $240$ to $600$ tasks/min. The advantage remains when only the admitted baseline tasks are counted, and the ablations confirm the importance of the graph interface, the hierarchical control, and the congestion information in the state. Without retraining, AeroGLIDER also keeps a $71.7\%$ success rate at an arrival rate $50\%$ beyond the training maximum and transfers to unseen network scales from $0.5\times$ to $2\times$ the training topology. Moreover, the worker needs only $4.8$--$5.5$~ms per decision on a CPU, well within the $50$-ms micro-slot. These results show that AeroGLIDER is a scalable and adaptive control framework for dynamic HAPS-UAV-IoT edge computing systems. Future work will extend AeroGLIDER to mobile UAVs, multi-HAPS
coordination, and distributed graph-policy training.

\section*{Acknowledgement}
This work was supported in part by the Natural Sciences and Engineering Research Council Canada (NSERC) Discovery and CREATE TRAVERSAL Programs, and in part by the Ontario Research fund-Research Excellence (ORF-RE) program under Project no. ORF-RE012-026.
\bibliographystyle{IEEEtran}
\bibliography{refs}

\end{document}